\documentclass[fleqn,10pt]{wlscirep}
\usepackage[utf8]{inputenc}
\usepackage[T1]{fontenc}
\usepackage{subfig}
\usepackage{cite}
\usepackage{amsmath,amssymb,amsfonts,color,colortbl,textgreek}
\usepackage{algorithmic}
\usepackage{graphicx}\usepackage{subcaption}
\usepackage{textcomp, url,hyperref}
\title{Analysis, Design, and Fabrication of a High-Gain Low-Profile Metasurface Antenna Using Direct Feeding of Sievenpiper's HIS}

\author[1]{Alireza Ghaneizadeh}
\author[2]{Sören F. Peik}
\author[3]{Martin Schneider}
\author[1,*]{Mojtaba Joodaki}
\affil[1]{School of Computer Science and Engineering, Constructor University, Bremen, 28759, Germany}
\affil[2]{Fakult\" {a}t Elektrotechnik und Informatik, Hochschule Bremen, Bremen, 28199, Germany}
\affil[3]{RF and Microwave Engineering Laboratory, University of Bremen, Bremen, 28334, Germany}

\affil[*]{mjoodaki@constructor.university}

\begin{abstract}
 High Impedance Surfaces (HIS) have recently shown the ability to support leaky waves, and to excite plasmonic and HIS resonance frequency modes for use as an antenna. In this paper, we analyzed, designed, and fabricated a true metasurface antenna (TMA) by directly feeding edge-located HIS cells through a microstrip feeding network. In contrast to other metasurface antennas that necessitate an external antenna to excite metasurfaces, our approach is inspired by the TMA design methodology that directly feeds the HIS cells rather than using it as a reflector. We developed a circuit model for the proposed structure and compared the results with those obtained from full-wave simulations. In addition, our further objective was to simplify the structure based on the working principle of the proposed antenna. This objective was achieved by converting square patches into parallel strip lines, leading to an aperture efficiency of about 77$\%$. This simplification also creates additional space to explore various resonant patterns on the top surface and the feeding network on the bottom surface of the TMA. Full-wave simulation results indicate that, despite the compact dimensions of the proposed array with 64 electrically small patch resonators  ($1.84\lambda\times1.84\lambda\times0.032\lambda$, where $\lambda$ is the free space wavelength at 6.0 GHz), it achieves a realized gain, HPBW (3dB beamwidth) of about 15.1 dBi and 28$^\circ$ respectively at 6.0 GHz. Finally, we constructed a prototype and conducted measurements to validate the design. Measured results demonstrate good agreement with simulation ones with a gain of about 13.5 (\(\pm\)0.5) dBi and a HPBW of 27$^\circ$ at 6.0 GHz. The proposed TMA is scaled to fit within the required dimensions for smart handheld devices at higher frequencies (e.g., 54 GHz), while maintaining high gain capability. The design's scalability, single-feed, and compact footprint make it optimal for diverse wireless communication systems, such as car-to-car (C2C) communications.
\end{abstract}
\begin{document}

\flushbottom
\maketitle
%
%
\thispagestyle{empty}

\section*{Introduction}

The demand for high-gain planar antennas has been steadily increasing in recent years due to their various applications in areas such as direct broadcasting satellite (DBS), radar and wireless power transfer\cite{Mousavi_k_2010,mosallaei_antenna_2004,alexandropoulos2023hybrid,sarabandi_design_2003,wu_simple_2023,nadi_multimode_2023,wang_pattern_2022,Bagheri_high_2023,Araghi_guided_2022}. Furthermore,  the emergence of 6G wireless communication systems has made it crucial to not only optimize antenna gain, compactness\cite{zheng_broadband_2024}, and ease of fabrication/integration onto planar platforms but also to enhance other critical antenna characteristics, such as aperture efficiency and beamwidth \cite{Ghanei_MEH_2024,Basar_wireless_2019,Babazadeh_circular_2018}. This is primarily due to the increasing number of antennas leading to energy and wave interference challenges, especially for short-range machine-to-machine (M2M) 6G wireless services.
Using an antenna producing the narrowest beams can reduce the interference \cite{yaman_vary_2019} and enhance the signal-to-noise ratio (SNR) by focusing the RF energy in a specific direction. This feature allows for a greater number of users to connect in today's high client-density wireless communication areas without causing interference in different spatial locations. Notably, every
three-decibel increase in planar antenna gain within the same
footprint area halves energy consumption. Therefore, narrow beamwidth antennas that exhibit maximum aperture efficiency (i.e., high-gain and compact radiators) will be demanded to meet the next generation of wireless technology requirements, addressing energy consumption and wave beam interference management.

To confine antenna radiation to half-space, a surface of a metal-backed dielectric layer or an electric wall is employed as the ground of the planar antenna, which can provide approximately hemisphere radiation coverage and shielding mechanisms \cite{guclu_direct_2011}. However, destructive interference arises when the distance between a horizontal dipole antenna and its parallel ground plane is reduced to less than a quarter of the wavelength, leading to decreased radiation efficiency\cite{luukkonen_high_2010}. 

In the late 1990s, a proposed solution to this limitation involved replacing traditional electric walls, which have a reflection phase $180^\circ$ out of phase, with artificial magnetic conductor (AMC) ground planes, referred to as metasurfaces \cite{GLYBOVSKI_metasurface_2016,Ghanei_general_2024,ghaneizadeh_compact_2020,Taghvaee_Overview_2024}. These AMC ground planes feature a nonconstant reflection phase ranging between $\pm180^\circ$ versus frequency \cite{balanis_evolution_2024,deo_thickness_2010}. This unique feature is in contrast to the perfect electric conductor (PEC) and perfect magnetic conductor (PMC) surfaces that maintain a $180^\circ$ and $0^\circ$ reflection phase, respectively \cite{chen_checkerboard_2015,balanis_evolution_2024,yang_reflection_2003}. The working principle of AMC is based on applying a reactive capacitive composite layer onto a metallic mirror. According to the transmission line theory, a perfect magnetic wall is equivalent to a surface in free-space positioned at a quarter-wavelength distance from a perfect electric conductor surface (i.e., $\lambda$/4 impedance inverter). This perfect magnetic wall has infinite surface impedance, leading to a reflection magnitude of unity and a phase of zero at its resonance frequency. A well-known example of the AMC structures is Sievenpiper's high-impedance surface (HIS), commonly referred to as the mushroom structure \cite{HIS_thesis_1999,hampel_mimo_2008,hampel_design_2007,hampel_sievenpiper_2008}. This structure has various applications within a limited frequency band, i.e., it can function as a reflector for low-profile dipole antennas\cite{soltani_multi_2020}, stop-band filters for surface waves, and reduce surface waves and edge diffractions over the ground plane \cite{yang_electromagnetic_2009,balanis_evolution_2024,luukkonen_effects_2009}. Additionally, the structure is utilized for positioning antennas close and parallel with an HIS ground plane \cite{Compact_thesis_2012,amiri_analysis_2016}.

An HIS configuration is typically a reactive composite thin layer that includes two elements: (i) a capacitive frequency characteristic layer, often formed by an array of square metal patches \cite{Radi_thin_2015,hampel_design_2007}. This layer is printed on (ii) a meta-backed dielectric substrate (which could be an electrically thin air substrate) with an inductive frequency characteristic \cite{Compact_thesis_2012,hampel_sievenpiper_2008}. Various patch shapes can be chosen for the top surface resonator of the HIS array, but all implementations fundamentally employ the same principle \cite{Radi_thin_2015}. It may include vertical vias connecting two metallic layers as the mushroom-textured surface, or it may not \cite{Goussetis_Tailoring_2006,feresidis_artificial_2005}. However, the vias are necessary to reduce surface waves in the substrate\cite{balanis_evolution_2024}. Furthermore, vias are crucial for oblique incidence applications, particularly in absorber applications \cite{Radi_thin_2015}. The authors in \cite{rajo_size_2007} proposed relocating the vias of the mushroom-type HIS from the center of the unit-cell to its edge to achieve a 20$\%$ reduction in its resonant frequency (see Fig.~1 in \cite{rajo_size_2007}). 

HIS architecture is engineered to modify the impedance boundary conditions of a surface, effectively creating a magnetic wall at its resonance frequency. These conditions facilitate the design of extremely thin planar antennas, allowing parallel electric currents to be positioned close to the HIS ground when the antenna's resonance frequency falls within the $\pm90^\circ$ reflection phase range of the HIS. Within this range, the HIS supports in-phase image currents aligned with the overlying antenna's currents parallel to the HIS \cite{balanis_evolution_2024,deo_thickness_2010}. For instance, the authors in \cite{yang_reflection_2003} used a mushroom-type HIS as the ground for a proximity transverse dipolar antenna to enhance its performance. The authors in \cite{mosallaei_antenna_2004} utilized a patch-FSS array on a metallic ground substrate to improve the bandwidth and miniaturization characteristics of the patch antenna. The advantages of using a bow-tie antenna on a mushroom-type HIS were also investigated in \cite{best_design_2008}. In a numerical study, an antenna was designed only with a mushroom-type HIS, which does not require a horizontal dipole antenna \cite{luukkonen_high_2010}. However, their approach involved the utilization of two feeding ports to excite two resonant surface modes: the plasmonic resonance mode \cite{rotman_plasma_1962,belov_strong_2003} and the HIS structure resonance mode \cite{luukkonen_high_2010}. A coaxial feed was implemented at the central via of a 3$\times$3 unit-cell array, and a voltage feed was applied in the dielectric gap between neighboring patches to excite these modes \cite{luukkonen_high_2010}, respectively. It is worth noting that the plasmonic resonance is generated by the current flowing through the vias, which functions similarly to a short monopole\cite{luukkonen_high_2010,antoniades_folded_2008}. Subsequently, the radiation properties of the mushroom-type HIS were experimentally explored and demonstrated in \cite{Karilainen_high_2011}. In this setup, the HIS was excited by a coaxial feed positioned below the central element of a 3$\times$3 unit-cell array to excite the plasmonic resonance mode, while a dipole antenna was placed on top of the HIS for the structure resonance mode (see Fig.~2 in \cite{Karilainen_high_2011}).
In \cite{fong_scalar_2010}, the authors introduced a holography technique to control the radiation of surface currents on a metasurface structure. This methodology facilitated the transformation of an input wave into a desired output wave by manipulating surface impedance. They utilized an Artificial Impedance Surface (AIS) comprising an array of sub-wavelength metallic patches on a metal-backed dielectric substrate. In this work, the modification of patch sizes enabled the engineering of the surface impedance pattern.

In \cite{vallecchi_low_2012}, an HIS made of “dogbone" elements was used to enhance the performance of a folded dipole antenna. In \cite{guclu_direct_2011}, the authors demonstrated the potential of directly feeding the same HIS layer as in \cite{vallecchi_low_2012}, showing that the HIS can operate as a main radiator without requiring an antenna to be mounted on top. Furthermore, in order to enhance the radiated gain and simplify the feeding of the structure in \cite{guclu_direct_2011}, a new feeding method for the HIS antenna was proposed in \cite{Almutawa_analyze_2018}. While this method reduces antenna thickness, it still employs four ports on two metal strips \cite{Almutawa_analyze_2018}, which are interconnected with the edges of the dogbones array. Similar techniques have been developed in\cite{guclu_high_2011,guclu_possible_2012,almutawa_ultrathin_2019,reazul_beamsteering_2020,reazul_small_2020,tanmoy_wide_2020}. As mentioned in \cite{Almutawa_analyze_2018}, this type of HIS antenna is a leaky-wave antenna characterized by a high attenuation constant, differing from standard leaky-wave antennas \cite{Almutawa_analyze_2018,allen_leaky_2004,jackson_leaky_2012,araghi_long_2024,Torabi_miniaturized_2023}. Additionally, the waveguide-fed metasurface antenna discussed in \cite{smith_analysis_2017,shlezinger_dynamic_2021,yoo_design_2022} uses cells that function as polarizable dipoles. These dipole cells facilitate the coupling of the waveguide mode to the radiation mode, as detailed in \cite{smith_analysis_2017,shlezinger_dynamic_2021}.

In \cite{badawe_true_2016}, the authors introduced the concept of a single-beam metasurface antenna with dense apertures comprising electrically small cross-shaped resonators with a pitch much shorter than half the wavelength, which is typical in traditional array antennas \cite{badawe_true_2016}. This revolutionary approach of the antenna design exposing a high aperture efficiency includes the resonators that are directly fed by metallic vias connected to a corporate feeding network with a single output terminal port, rather than using an external antenna\cite{koohestani_ultra_2021} or waveguide feeding technique \cite{smith_analysis_2017}. In addition, the authors in \cite{zhang_low_2024} proposed a periodic metasurface structure incorporating a square patch enclosed by a squared ring designed for energy harvesting and polarization detection applications. In comparison to metasurface-loaded antennas that require an external antenna to excite the metasurfaces\cite{koohestani_ultra_2021}, true metasurface antenna (TMA) structures which are self-excited and do not require an external antenna, effectively address issues such as shadow effects in reflect arrays, larger sizes, grating lobes, low aperture efficiency, and difficulties related to phase compensation \cite{badawe_true_2016,rajabalipanah_highly_2020}.

Although HISs are commonly used to enhance the overall radiation performance of existing antennas, the authors in \cite{Mu_self_2024} and \cite{chang_tailless_2024} demonstrated that square and rectangular patches can operate as independent antenna elements. For example, in \cite{Mu_self_2024}, the designed metasurface structures are comprised of subarrays, each of which is constructed by connecting groups of four patch cells through a power divider. All subarrays were then excited simultaneously via their individual ports. In particular, the contributions of this article can be summarized as follows:
\begin{itemize}
\item We exploit Sievenpiper’s HIS by directly feeding squared patches to analyse patch metasurface antenna. To do this, we analyzed the circuit model and radiative properties of the surface current, magnetic field and electric field distributions. Considering the extensive efforts made by researchers since 1999\cite{Sievenpiper_high_1999} to develop HIS-based structures for various applications, this metasurface antenna fed by a designed microstrip feeding pattern may open new avenues in these fields.

\item According to the working principle of the proposed single port metasurface antenna, we further simplified the design by removing the vertical slots on each patch and converting them into long continuous stubs loaded with vias, which are connected to the feeding network on the ground layer. This simplification approach has the potential to reduce the number of vias while maintaining comparable performance to the previous resonator style, thereby allowing for increased space to implement a feeding network pattern.

\end{itemize}
In the following sections, we will demonstrate how we utilized a microstrip feeding network to excite all edge-located via HIS. The key difference between high-impedance mushroom surfaces and our TMA cells lies in the configuration of the vias. In high-impedance mushroom surfaces, the vias connect the top surface of the HIS to the ground. In contrast, our TMA design features vias that are isolated from the ground and connected to the microstrip feeding network. In the TMA, the ground is shared between the top patches and the microstrip feeding network pattern on the bottom layer.


\section*{Metasurface Antenna Design And Simulation}

The structure of designed TMAs involves three metallic patterns and two dielectric layers. As already mentioned, the top layer consists of the Frequency Selective Surface (FSS), while the bottom layer constitutes the feeding network. The middle layers are a common ground between the FSS and the feeding network. Moreover, metal vias establish connections from the feeding network to each FSS, simultaneously providing isolation of the shared ground metal layer from these metal vias. Inspired by true metasurface antennas \cite{badawe_true_2016,rajabalipanah_highly_2020,Nadi_flexible_2020}, all meta-radiators are linked to a feed point network in a manner that ensures their impedance matches the feed impedance, resulting in the antenna's optimum bandwidth and gain.

\subsection*{Lumped Circuit Modeling of TMA's Unit-Cell}
Full-wave numerical EM software like HFSS and CST are commonly employed for TMA analysis\cite{Ghaneizadeh_circular_2024}, which is a complex and time-consuming task. In addition, the outcomes of these simulations are affected by different parameters, e.g., feed position, substrate thickness, and mutual coupling between TMA cells \cite{ghaneizadeh_analysis_2020}. In our previous works on metasurface energy harvesters, we modeled a patch unit-cell with four via \cite{ghaneizadeh_analysis_2020} and a complementary quad split ring resonator (CQSRR) cell \cite{ghaneizadeh_extremely_2020} using an equivalent circuit. The circuit model presented in Fig.~1d has been adapted from \cite{ghaneizadeh_analysis_2020,ghaneizadeh_extremely_2020,koohestani_ultra_2021} for the patch cell with a single via. Analyzing the TMA's cell with an equivalent circuit model provides insights into its physical characteristics and estimates its electromagnetic response. Despite its simplicity, an equivalent circuit approach helps to understand the operational mechanism of a TMA.

Generally, an array of metallic square patches with sub-wavelength periodicity operates as a dipole-type FSS array with capacitive properties when interacting with plane waves\cite{ghaneizadeh_analysis_2020}. Basically, both inductance and capacitance are required for a resonant structure in the path of the electromagnetic wave. Consequently, placing the patches on a grounded substrate creates a resonant structure since the dielectric thin substrate with the metallic ground plane has inductive properties \cite{ghaneizadeh_analysis_2020}. The equivalent circuit of the proposed unit-cell is demonstrated in Fig.~1. For a better illustration, the position of each lumped circuit element on the TMA's cell is specified in Fig.~1a-c. Fig.~1e displays the scattering parameters, including the reflection and transmission coefficients, obtained from the ADS circuit simulator for the circuit model of the TMA's unit-cell. For the sake of clarity, these results are compared with those from the full-wave simulator in CST. The responses of both the circuit and the full-wave simulations in Fig.~1e exhibit a good similarity. It is noted that port one in Fig.~1 is employed for a normal incident wave impedance in air or vacuum with an approximated value of $\mathit{Z}_0$ = 377 $\Omega$, and port two is used for the load of TMA cell with $R_L$ = 70 $\Omega$.

\begin{figure}[!t]
    \centering
    \subfloat[]{%
        \includegraphics[width=0.2\linewidth]{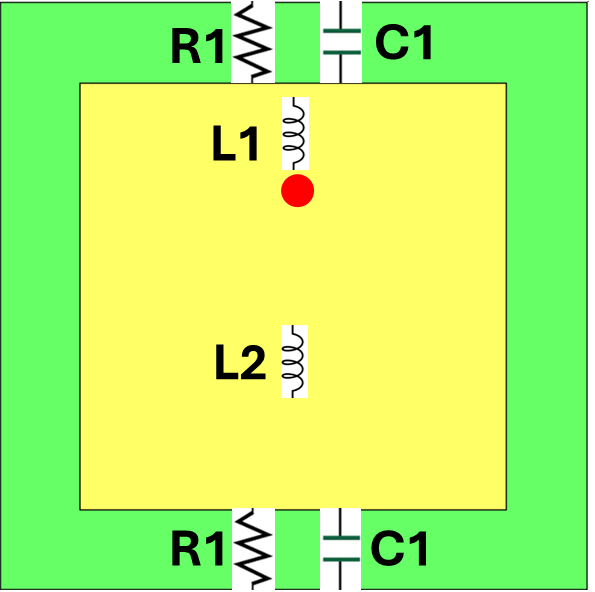}
    }
    \hspace{0.03\linewidth}
    \subfloat[]{%
        \includegraphics[width=0.2\linewidth]{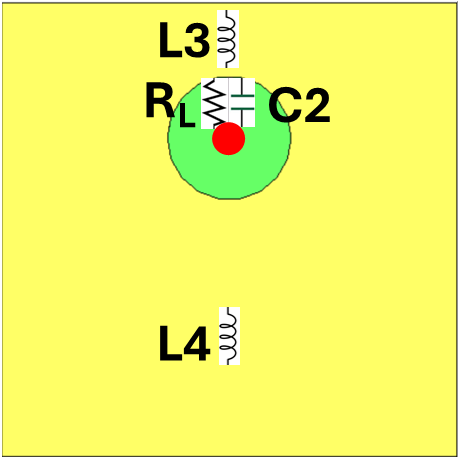}
    }
    \hspace{0.03\linewidth}
    \subfloat[]{%
        \includegraphics[width=0.3\linewidth]{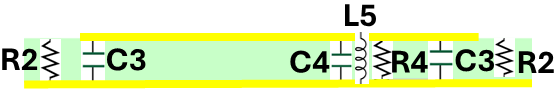}
    }
    \newline
    \subfloat[]{%
        \includegraphics[width=0.3\linewidth]{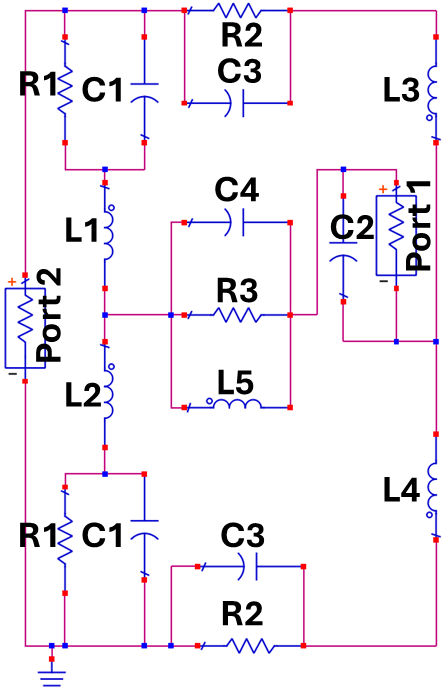}
    }
    \hspace{0.05\linewidth}
    \subfloat[]{%
        \includegraphics[width=0.34\linewidth]{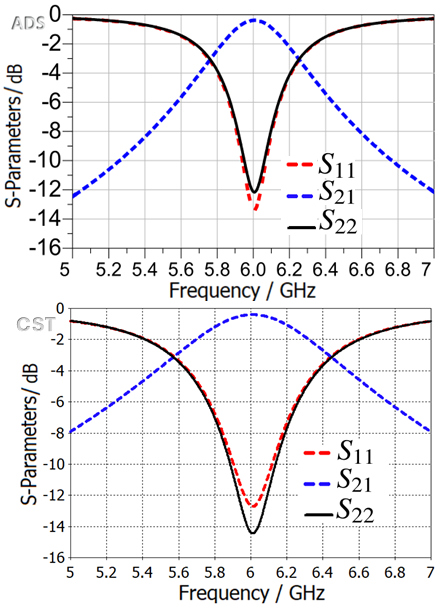}
    }
    \caption{(a) Front, (b) back, and (c) side views of the radiating cell, (d) an equivalent circuit model of the patch unit-cell, and (e) the simulation results of unit-cell using CST and its equivalent circuit model using ADS. [Noted that the schematic views of cells in (a)-(c) are in a different scale for a better illustration. Port 1, with an impedance of $\mathit{Z}_L$ = 70 $\Omega$, represents the load of the unit-cell, while port 2, with an impedance of $\mathit{Z}_0$ = 377 $\Omega$, represents the free-space wave impedance.]}
\end{figure}

An initial estimate was determined for some of the lumped components, and then the equivalent circuit parameters were optimized by ADS circuit simulation software \cite{costa_efficient_2012,koohestani_ultra_2021,yang_polarization_2021}. Sievenpiper \textit{et al.} introduced a calculation method for analysing the mushroom-like HIS with a central via \cite{HIS_thesis_1999,liu_effects_2024}. For example, the equivalent capacitance $C_0$ generated from the top patches between two adjacent unit-cells of the HIS was estimated by eq.~(1) \cite{liu_effects_2024}.

\begin{equation}
C_0 = \frac{L(\varepsilon_0 + \varepsilon_1)}{\pi} \cosh^{-1}\left(\frac{L + 2s}{2s}\right)
\end{equation}
where $\varepsilon_0$ is the dielectric constant of vacuum, $\varepsilon_1$ is the dielectric constant of the HIS substrate, \textit{L} is the side length of the square patch, and \textit{s} represents the distance between the patch and the edge of the unit cell (Fig.~2).  

The equivalent inductance ($L_0$), resulting from the top metal patches and the bottom metal ground, was also derived from eq.~(2)\cite{liu_effects_2024}.

\begin{equation}
L_0 = \mu t
\end{equation}
where $\mu$ is the permeability of the HIS substrate, and \textit{t} is its thickness (see Fig.~3 in \cite{liu_effects_2024}).

Additionally, the inductance of the metallic via ($L_v$) was estimated using eq.~(3) \cite{liu_effects_2024}. 

\begin{equation}
L_v = \frac{\mu t}{2\pi} \ln \frac{L + 2s}{2r} = 0.16 \mu t \left( \ln \left( \frac{L + 2s}{r} \right) - 0.69 \right)
\end{equation}
where \textit{r} reperesents the outer radius of the metallic via.

Since our TMA cells are similar to the edge located via HIS except for the port, we have used these approximations for the initial values of our circuit model and then optimized them using ADS simulation software. Additionally, the static capacitance of an ideal parallel-plate capacitor ($\
C = \frac{\varepsilon_0 \varepsilon_r A}{t}
$) was used to estimate the equivalent capacitance formed between parallel metal plates of the top and bottom layers \cite{liu_effects_2024}, where $\varepsilon_0$ is the permittivity of free space and  $\varepsilon_r$ is the relative permittivity. The average area of the bottom and top metal layers of TMA's cell was considered for calculating area $A$ \cite{ghaneizadeh_analysis_2020}. Since the electric fields concentrated between adjacent patches are much less than those concentrated between the top and bottom layers, the value of capacitance $C_1$ is much smaller than the capacitances of the substrate's top and bottom metal layers \cite{ghaneizadeh_analysis_2020}. It should be noted that the capacitance $C_0^{\text{patch}}$ formed within narrow gaps between the edges of adjacent square patches in a freestanding configuration was referred to as the unloaded capacitance. This unloaded capacitance describes the response of the periodic squared patches to a normal incident wave in a freestanding configuration and was estimated as follows \cite{costa_closed_2012}:



\begin{equation}
C_0^{\text{patch}} = \frac{2P\varepsilon_0}{\pi} \ln \left( \frac{1}{\sin\left( \frac{\pi [P - L]}{2P} \right)} \right)
\end{equation}
where \textit{P} is the periodicity of the unit-cell.

The ohmic losses of TMA's 35 µm thick copper layer can be ignored in the microwave range. However, a parallel resistor is included with each lossless capacitor to account for dielectric losses. The final values for the equivalent circuit elements are determined using a random optimization algorithm in ADS simulator as follows: $R_L$ = 70 $\Omega$, $Z_0$ = 377 $\Omega$, $C_1$ = 1.3 pF, $R_1$ = 31.05 k$\Omega$, $L_1$ = 0.51 nH, $C_2$ = 2.16 pF, $R_2$ = 1.31 k$\Omega$, $L_2$ = 0.1 nH, $C_3$ = 11.95 pF, $L_3$ = 11.5 nH, $R_3$ = 655.5 $\Omega$, $C_4$ = 5.6 pF, $L_4$ = 0.99 nH, $L_5$ = 2.19 nH. Although the initial estimation values are different from the values obtained after optimization with ADS software, they are suitable as initial estimates considering the complexity of the analytical equations for these elements. It is worth mentioning that the full-wave simulation is used for our final TMA design.


\subsection*{Unit-cell design}

The electrically small resonator of Sievenpiper's HIS has been adapted to serve as meta-radiator of TMA, as shown in Fig.~2. As already mentioned, the periodic arrangement of the symmetric patch connected to the grounded substrate with edge-located via is utilized to create an HIS (see Fig.~1b in \cite{rajo_size_2007}), which has different applications, such as enhancing the radiation performance of planar antennas.
\begin{figure}[t]
    \centering
    \subfloat[]{%
        \includegraphics[width=0.22\linewidth]{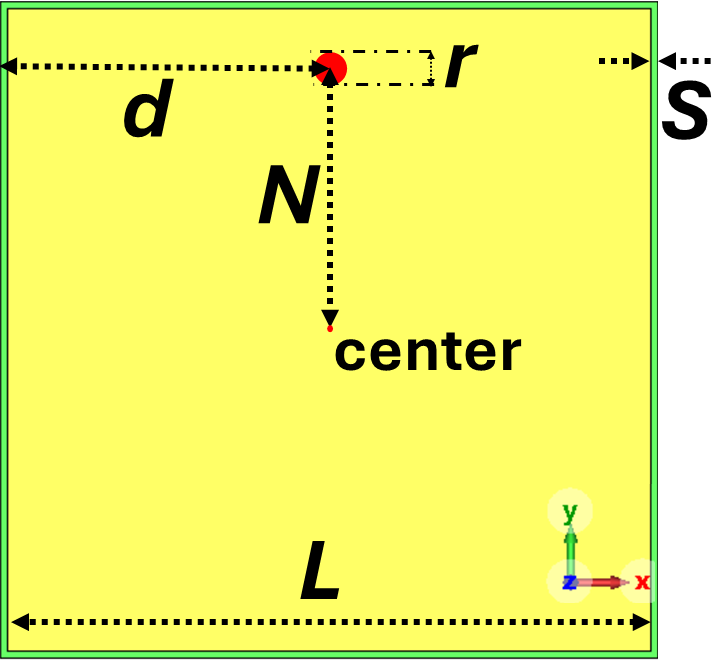}
    }
    \hspace{0.05\linewidth}
    \subfloat[]{%
        \includegraphics[width=0.2\linewidth]{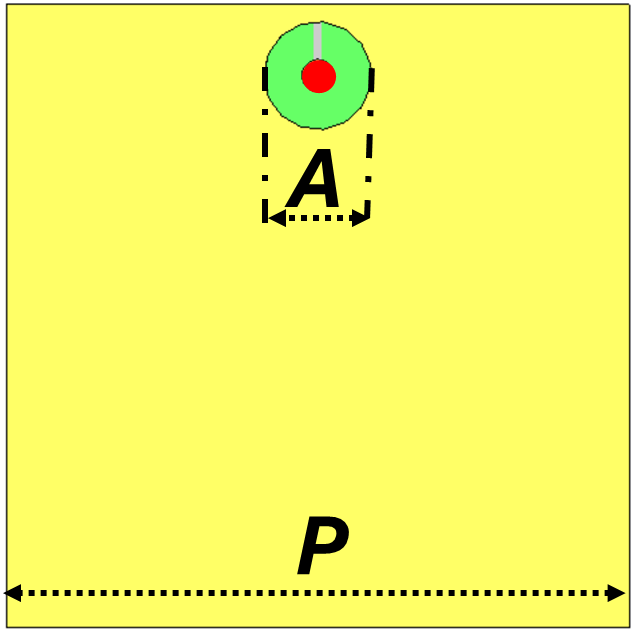}
    }
 \hspace{0.05\linewidth}
     \subfloat[]{%
        \includegraphics[width=0.3\linewidth]{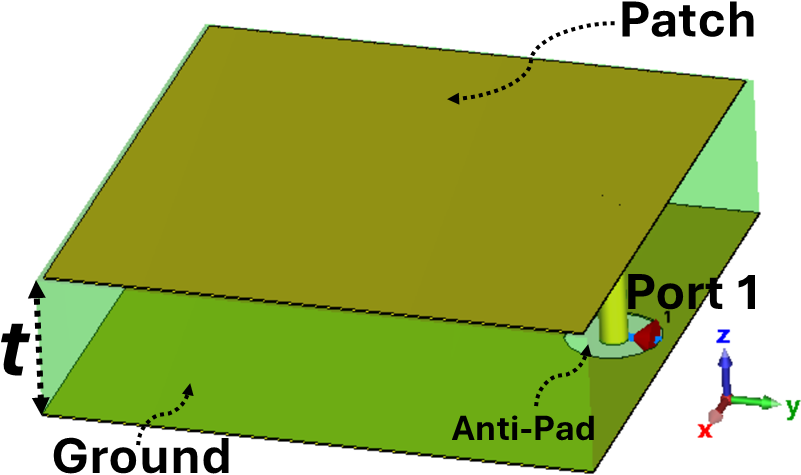}
    }
    \caption{A schematic view of the TMA cell with dimensions of \textit{L} = 11.28 mm, \textit{s} = 0.125 mm, \textit{N} = 4.45 mm, \textit{A} = 2 mm, \textit{P} = 2\textit{d} = 11.53 mm, \textit{t} = 0.813 mm, \textit{r} = 0.6 mm (a) top, (b) bottom, and (c) perspective view. [Noted that the thickness of the cell in (c) is in a different scale for a better illustration.]}
\end{figure}

The square metallic patches are etched on one side of the substrate of Rogers Ro4003 with a relative permittivity ($\varepsilon_r$) of 3.26, loss tangent ($\tan \delta$) of 0.0027, and a substrate thickness of 0.813 mm, while the opposite side serves as the metallic ground plane. In addition, a circular antipad with a 1 mm radius is etched around the via hole on the ground plane to isolate the metallic via from the ground. To achieve a minimal reflection coefficient at 6.0 GHz, the optimized geometric dimensions of the cell are determined as $L = 11.28$ mm, $S = 0.25$ mm for a copper thickness of 35 $\mu$m.

The behavior of the electrically small meta-radiators is studied by investigating the unit-cells, which are periodically positioned in the x and y directions and illuminated by a normal incident wave. For this purpose, a TMA's cell is located at the waveguide's center in the x-y plane\cite{badawe_true_2016}. In order to excite the TMA using a transverse electromagnetic (TEM) mode in the z-direction utilizing the CST commercial software, perfect magnetic walls are positioned in the y-z plane, and perfect electric walls are positioned in the x-z plane of the waveguide \cite{badawe_true_2016}. Two waveguide ports, labeled Port 2 and Port 3, are located on the top and bottom sides of the waveguide, respectively, to excite the unit-cell with a plane wave from top surface of the unit-cell. In addition, a lumped port, labeled Port 1, with an impedance of 70 $\Omega$ is positioned between the metallic via and the ground surface of the unit-cell to provide excitation from its termination. 

Using unit-cell boundary conditions in the Floquet analysis of CST Microwave Studio, the radiation profile of the TMA cell shows a symmetric and normal propagation along the broadside direction (Fig.~3a). The reflection coefficients $S_{11}$ and $S_{22}$ of the infinite array of designed radiating cells are shown in Fig.~3b. In addition, Fig.~3b shows that the impedance of the top surface of TMA is matched with the impedance of the incident normal radiation wave (i.e., $S_{22}$ in Fig.~3). These results indicate that the antenna surface behaves as a metasurface at its operating frequency because this antenna surface can be represented as an equivalent surface characterized by nearly equal relative permeability and permittivity \cite{badawe_true_2016}. The location of metallic via has been optimized using the transformer model of the cell introduced in \cite{ghaneizadeh_extremely_2020} and is shown in Fig.~3c. Due to limitations in TMA fabrication and the positioning of the anti-pad near the edge of the cell, as shown in Fig.~2b, we have set the maximum value of $\textit{N}$ = 4.45 mm. As a result, the via is situated approximately 4.5 mm from the center, providing a reflection coefficient of less than -10 dB.

\begin{figure}[t]
    \centering
    \subfloat[]{%
        \includegraphics[width=0.27\linewidth]{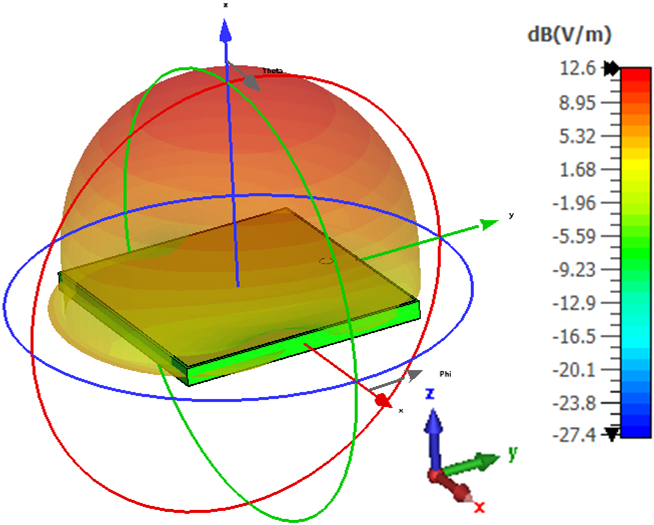}
    }
      \hspace{0.05\linewidth}
    \subfloat[]{%
        \includegraphics[width=0.25\linewidth]{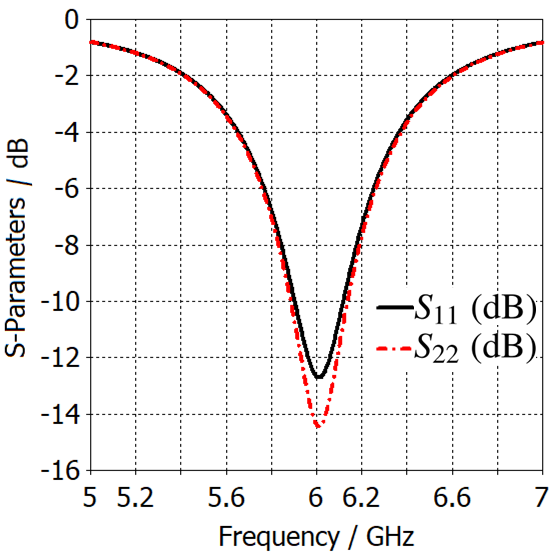}
    }
      \hspace{0.05\linewidth}
        \subfloat[]{%
            \includegraphics[width=0.29\linewidth]{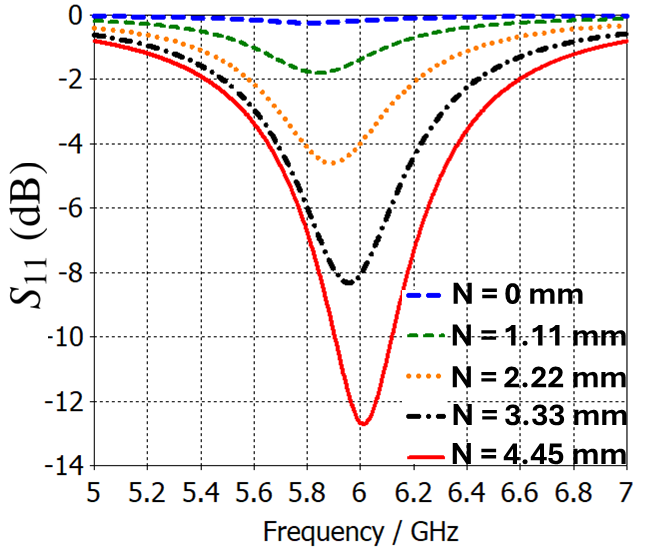}
        }
    
    \caption{(a) Unit-cell radiation pattern, (b) scattering parameters from the lumped port 1 ($S_{11}$) located at the load of the cell and from the waveguide port 2 ($S_{22}$) located at the waveguide incident port for illuminating the top surface of the radiating cells with a TEM mode (377 $\Omega$), and (c) the dependency of $S_{11}$ (dB) on the distance of the via-location between the center of the cell  ($N=0$ mm) and maximum available distance of the via location from the center ($N=4.45$ mm).}
\end{figure}


Fig.~4 demonstrates the distribution of a unit-cell's surface current strength, as well as the electric and magnetic field strengths in an infinite array, illuminated by a normal incident plane wave with y-polarization at the resonance frequency of 6.0 GHz for different relative phases of incident plane wave. The electric field is concentrated around the dielectric area between the edges of the square patch, which are parallel to the x-axis.  It is also observed that the electric field in the middle line of the cell, parallel to the x-axis, reaches its minimum. Furthermore, these results indicate that when an incident wave with y-polarization interacts with the unit-cell surface, the electric field almost vanishes in the insulating space between the edges parallel to the y-axis, effectively converting the patch array into a strip array \cite{clavijo_design_2003,luukkonen_efficient_2008,luukkonen_simple_2008}. Fig.~5 shows a TMA array of 8 strip lines with surface current distribution on the TMA at 6.0 GHz. The following section compares the simulation results for radiation gain and $S_{11}$ in both simplified strip and square patch configurations, showing very similar results (Fig.~9). In \cite{clavijo_design_2003}, the simplification method (i.e., transforming the patch to strip line, Fig.~5 in \cite{clavijo_design_2003}) is presented as a part of the design methodology for Sievenpiper's HIS. In addition, the process of fabricating a subwavelength scale Au nanostrip-array metasurface with 1400 nm-wide strips has been investigated in \cite{Park_dynamic_2017} for dynamic controlling of reflection phase and polarization wave at the operation frequency around 50 THz (see Fig.~2d in \cite{Park_dynamic_2017}).

\begin{figure}[!t]
    \centering
    \subfloat[]{%
        \includegraphics[width=0.46\linewidth]{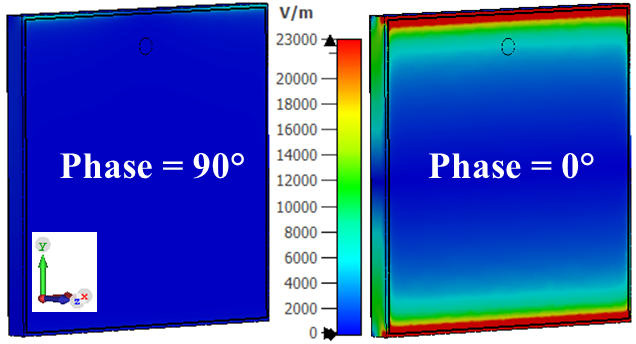}
    }
    \hspace{0.05\linewidth}
    \subfloat[]{%
        \includegraphics[width=0.47\linewidth]{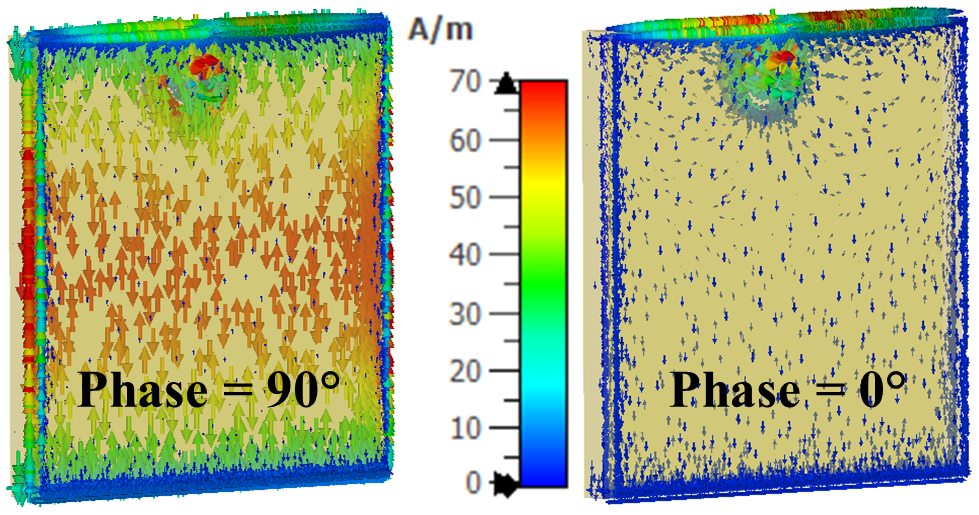}
    }
    \newline
  \subfloat[]{%
        \includegraphics[width=0.9\linewidth]{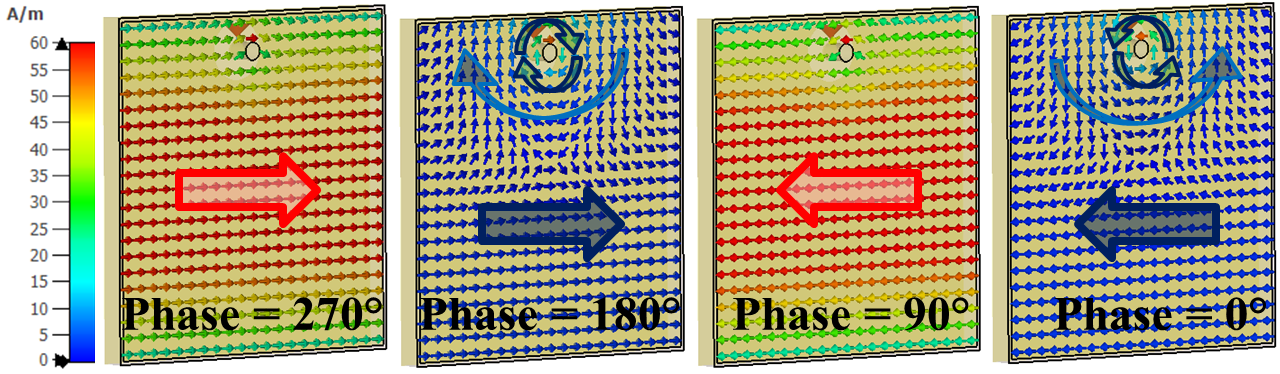}
    }

\caption{The distributions of the (a) electric field strength, (b) surface current strength and (c) magnetic field strength on the periodic unit-cells excited by a normal incident plane wave for different relative phases at the resonance frequency of 6.0 GHz.}
\end{figure}

\begin{figure}[!t]
    \centering
    \subfloat[]{%
        \includegraphics[width=0.3\linewidth]{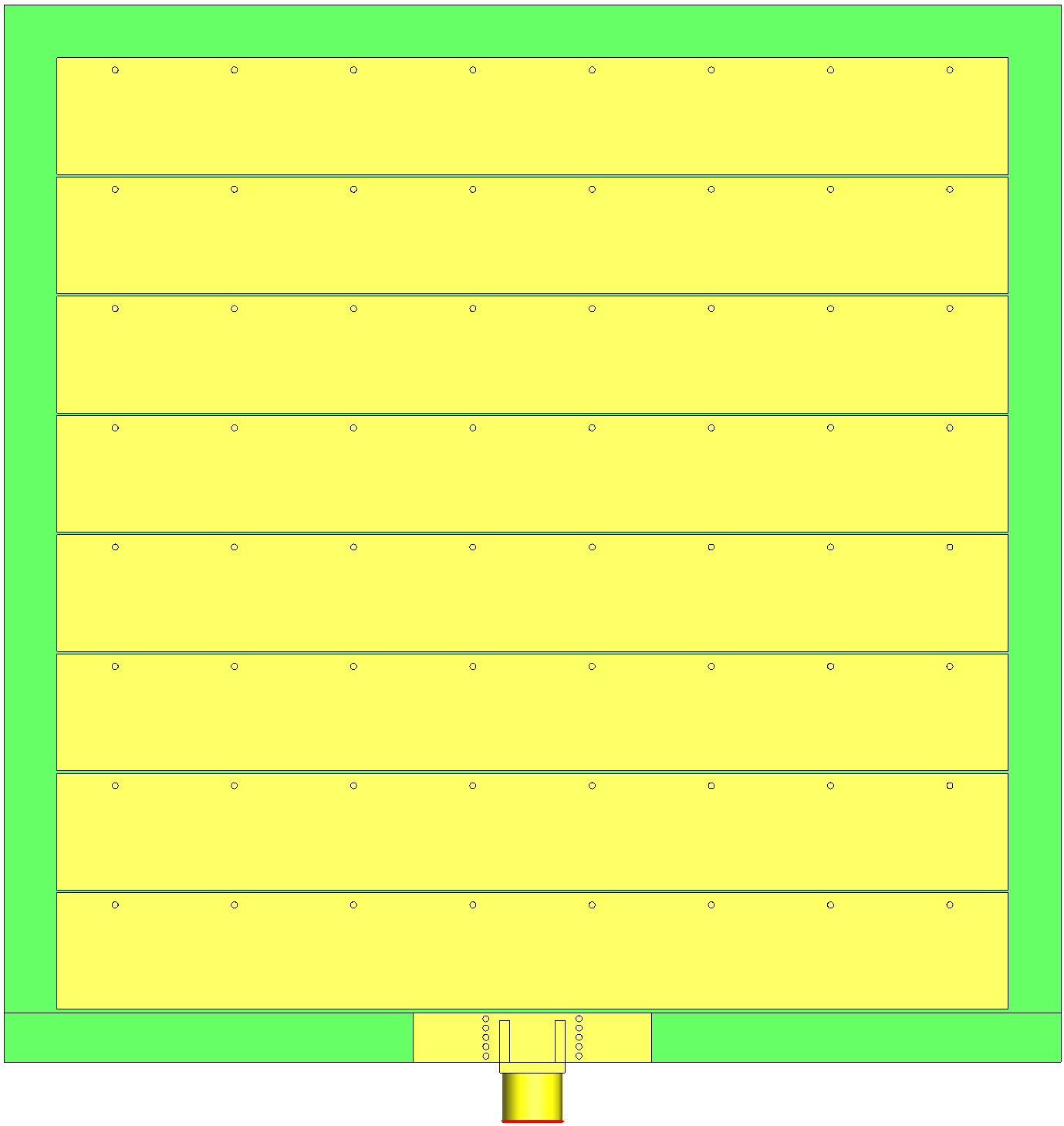}
    }
    \hspace{0.05\linewidth}
    \subfloat[]{%
        \includegraphics[width=0.34\linewidth]{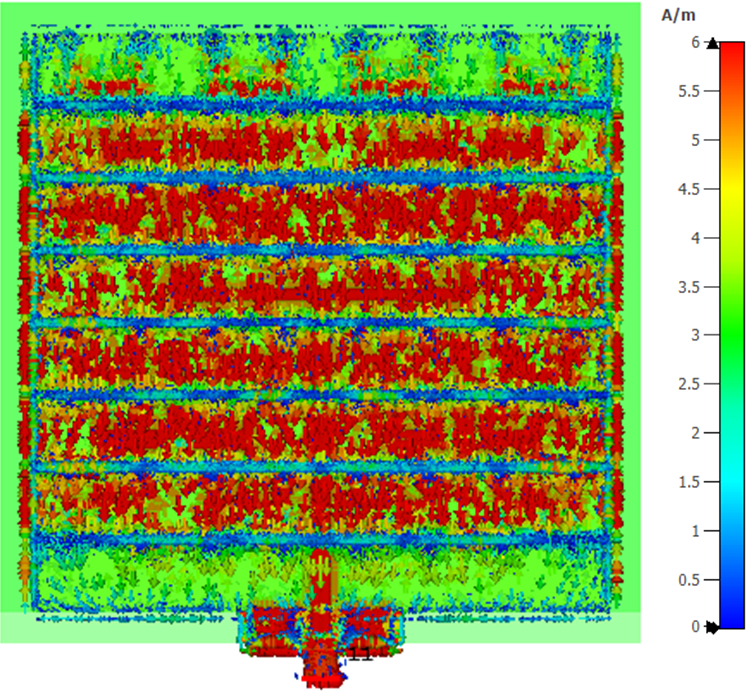}
    }
 
   \caption{ (a) The front view of TMA with eight strip line array, and (b) the distribution of surface current on the parallel strip TMA array at 6.0 GHz. (Noted that the square patch array is reduced to a strip array with metalizing just vertical slots in the patch cell, as shown in Fig.~2.)}
\end{figure}
The authors in \cite{ghaneizadeh_analysis_2020} show that the distance between the minimum and maximum electric field distribution is approximately $\lambda_g$/4 if the substrate thickness of the square patch cell and the ratio of $\textit{L}$/$\textit{P}$ is about 0.004$\lambda_g$ and 0.98, respectively, where $\lambda_g$ is the guided wavelength at the resonant frequency. Fig.~6 presents an analysis of how variations in dielectric substrate thickness affect the resonance frequency of unit-cells. The thickness ranges from 0.003$\lambda_g$ to 0.2$\lambda_g$, showing that the cell dimensions are approximately $\lambda_g/2$ when the substrate thickness is 0.003$\lambda_g$. This approach of metasurface antenna design simplifies the analysis process compared to conventional array antennas, focusing on implementing an appropriate feeding network after designing the TMA's cells.

\begin{figure}[t]
    \centering
    \subfloat{%
        \includegraphics[width=0.5\linewidth]{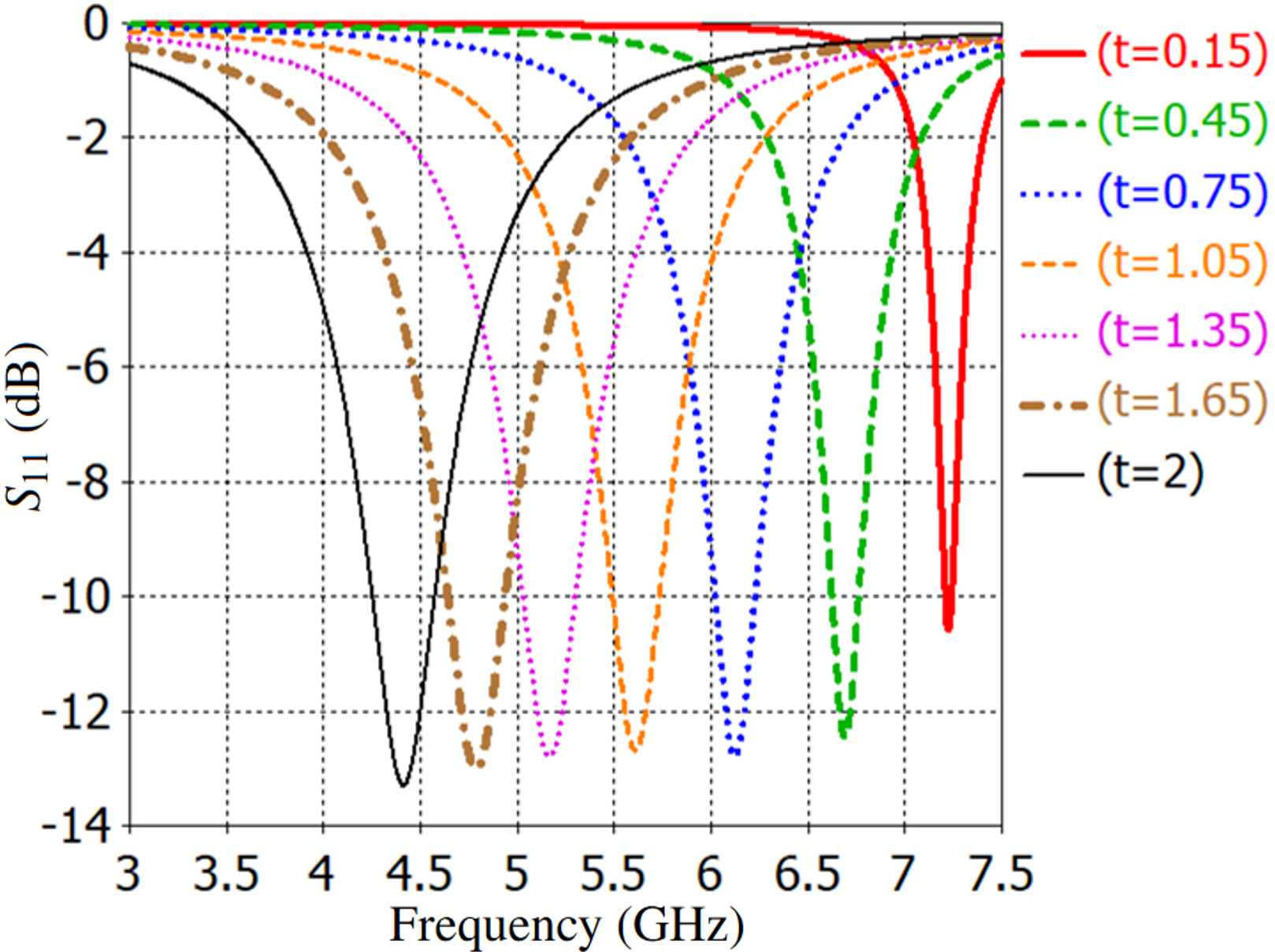}
    }
      \caption{Dependence of the $S_{11}$ on the thickness [$t$ (mm)] of the substrate when TMA is illuminated by a normal incident plane wave.}

\end{figure}

\subsection*{Simulation of the proposed TMA structure }

Unlike a conventional array antenna, a single cell of a metasurface antenna is not efficient individually \cite{badawe_true_2016}. However, when the elements
of a TMA are positioned close together, they
form an effective electromagnetic radiator\cite{rajabalipanah_highly_2020}. This behavior results from the appropriate inter-element coupling, which significantly influences the impedance matching of each cell to its designated feed port.

Fig.~7 illustrates the proposed TMA, consisting of an 8$\times$8 array of radiating cells on a platform of 92.24 mm $\times$ 92.24 mm, operating at a frequency of 6.0 GHz. Additionally, a 5mm border of Rogers Ro4003 substrate (green color) is used around the TMA for mounting plastic screws. The distance between the patches of the TMA is 0.25 mm, which is significantly smaller than the typical half-wavelength spacing between elements in conventional array antennas. This means that a larger number of meta-radiators can fit within the same space.

\begin{figure}[!t]
    \centering
    \subfloat[]{%
        \includegraphics[width=0.3\linewidth]{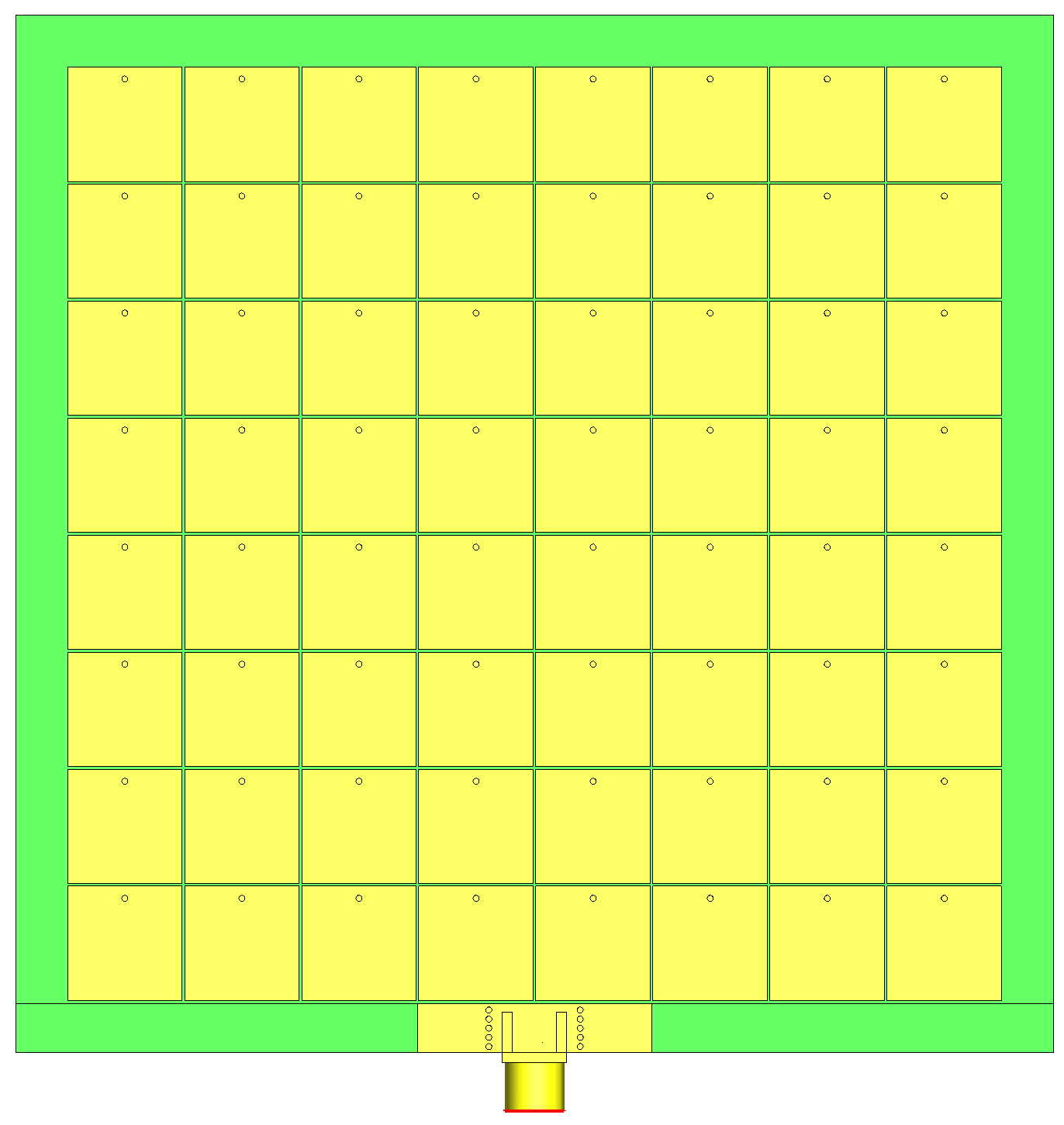}
    }
    \hspace{0.05\linewidth}
    \subfloat[]{%
        \includegraphics[width=0.3\linewidth]{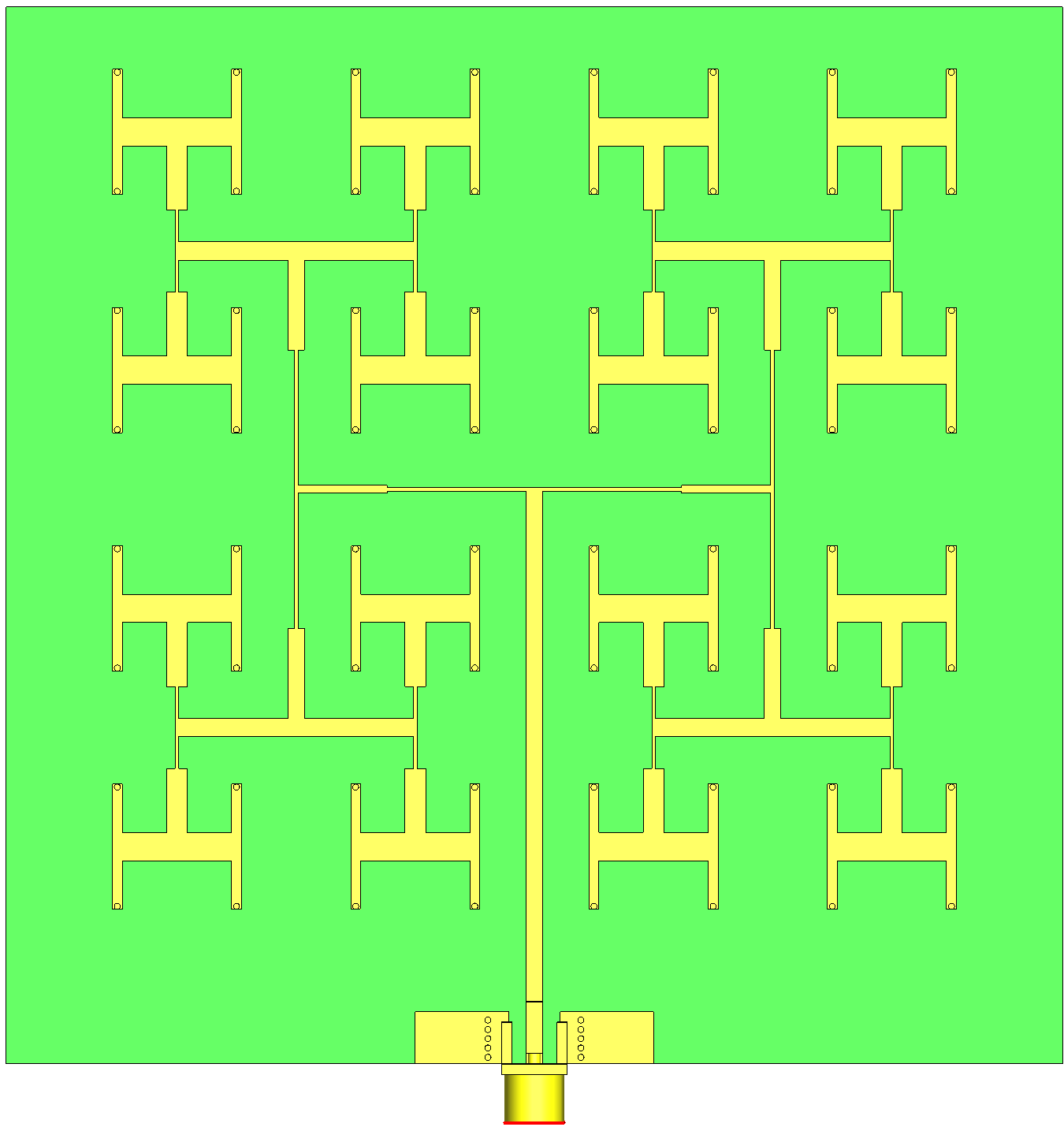}
    }

    \caption{ 
    (a) The front view of proposed TMA  with 64 square patches and (b) the back view of TMA including a corporate feeding pattern network of a 64-cell array [Noted that a 5mm border of Rogers Ro4003 substrate (green color) is used around the TMA for mounting plastic screws.]}
\end{figure}

Traditional array antennas typically have enough space to include a feeding network within a single substrate \cite{badawe_true_2016}. However, due to the proximity of TMA's cells, TMA design requires at least two dielectric substrates to implement the feeding network pattern. Therefore, we incorporate two Rogers Ro4003 substrates with a thickness of 0.813 mm and a metallic plane in the middle layer. This metallic plane, with a thickness of 70 $\mu$m, serves as a shared ground plane for the meta-radiators and the microstrip-type feeding pattern. The thickness of the metallic layer of the top and bottom layer of TMA is 35 $\mu$m. The sub-wavelength radiating cells are squared patches which are placed on the top surface and are connected to a SubMiniature version A (SMA) 50 $\Omega$ point through a feeding network pattern located below the ground plane (Fig.~7). The radiating elements are excited through metallic vias with a selected diameter of 0.6 mm to align with our fabrication constraints. It is worth mentioning that during fabrication, the metallic vias protrude outward about 0.7 mm from both the top and bottom layers of the TMA, which is included in the full-wave simulations. To achieve maximum gain in the boresight direction, equal lengths from the SMA feed point to each unit-cell are used to excite all meta-radiators simultaneously \cite{badawe_true_2016}. This configuration results in a nearly uniform distribution of surface currents across the unit-cells, except for the edge cells (Fig.~9). It should be noted that mutual coupling effects between unit-cells in a finite array and the feeding network impact the radiation pattern. 

In order to numerically investigate the impact of substrate thickness on impedance matching, we reduced the substrate thickness from 0.813 mm to 0.7 mm, as demonstrated in Fig.~8. Simulation results showed that this reduction caused a slight shift in the resonance frequency\cite{ghaneizadeh_design_2019}. The antenna's matching circuit design includes three quarter-wavelength impedance transformers. As shown in Fig.~6, the reduction in dielectric thickness of TMA's cells led to an increase in the resonance frequency of the array. Despite this, the length and width of the feeding circuit traces remained unchanged, leading to a deterioration in $S_{11}$ as shown in Fig.~8.

\begin{figure}[!t]
    \centering
    \subfloat{%
        \includegraphics[width=0.45\linewidth]{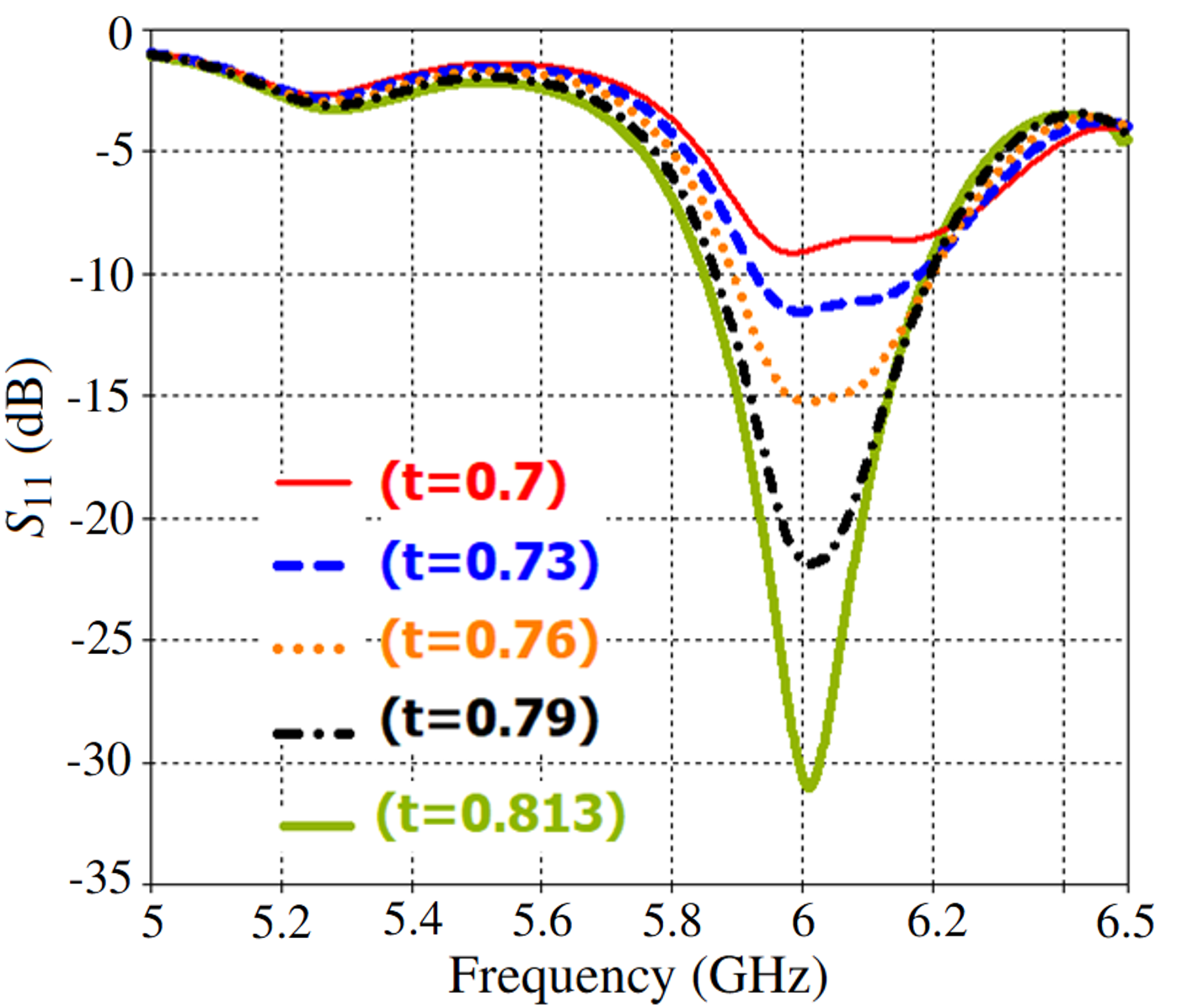}
    }
\caption{
The simulated results of $S_{11}$ variations with different thicknesses of the top and bottom dielectric substrates, ranging from 0.7 mm to 0.813 mm.}

\end{figure}

Fig.~10 presents the full-wave simulation results of the three-dimensional radiation pattern for the TMA with both squared-patch and strip-line elements. The $S_{11}$ of both arrays at the feeding port is approximately the same. However, the strip-line TMA shows slight improvements in the realized gain due to a more uniform distribution of surface currents in each cell leading to a better impedance matching. As shown in Figs. 5 and 8, the edge cells in the strip-line TMA exhibit nearly equal currents compared to the adjacent cells in the squared-patch TMA.

For better demonstration, Fig.~11 shows the simulation results of the $S_{11}$ for the strip-line cells, which are the same size as those in Fig.~2, except that only two vertical slots are metallized. In this figure, the inter-via distance along the x-axis is varied within an infinite strip-line array. This simplification approach allows us to decrease the number of vias by increasing the parameter $d$ in Fig.~2 while maintaining performance comparable to the previous resonator design. In fact, one of the challenges with the TMA design is the close spacing of feeding vias, which complicates the implementation of the feeding network due to limited space. However, by employing the strip line technique and increasing the parameter $d$ from its original value, as shown in Fig.~2, we can achieve relatively effective impedance matching while also providing more space for implementing the feeding pattern with fewer vias in the same footprint area of the TMA.

\begin{figure}[!t]
    \centering
    \subfloat[]{%
        \includegraphics[width=0.345\linewidth]{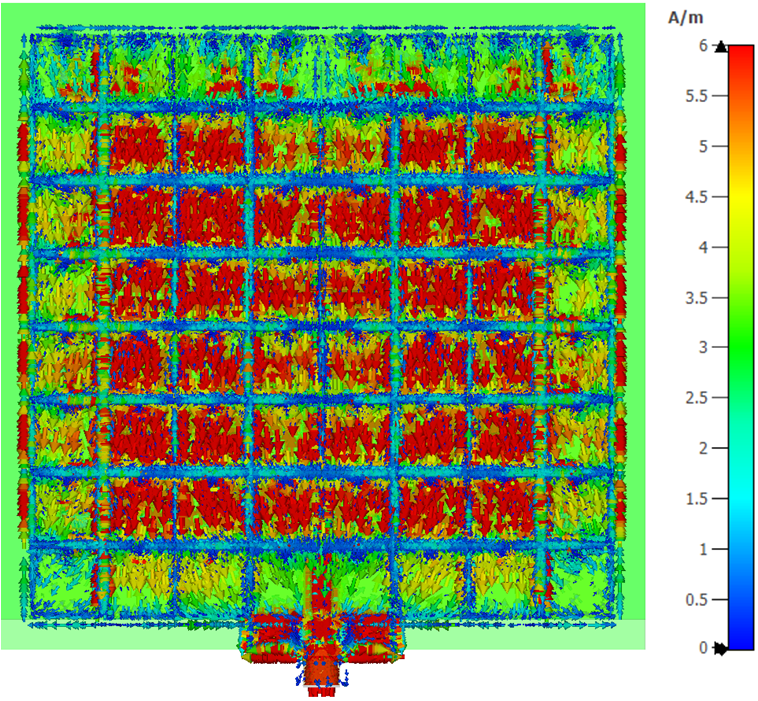}
    }
    \hspace{0.05\linewidth}
    \subfloat[]{%
        \includegraphics[width=0.3\linewidth]{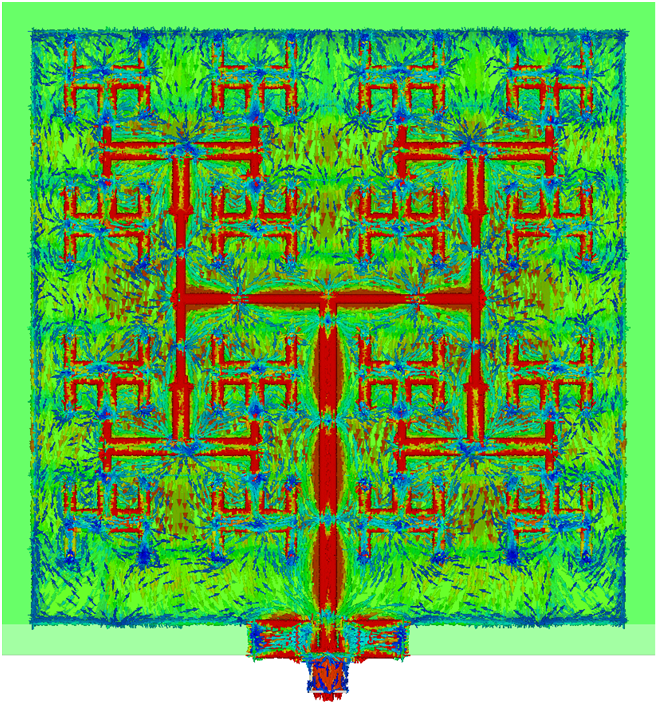}
    }
   
\caption{Surface current distribution on (a) the squared patch cells of TMA and (b) the microstrip feeding network pattern at 6.0 GHz.}

\end{figure}

To avoid fabrication complications associated with narrow microstrip line widths in the feed network, we use a load impedance of 70 $\Omega$ for each radiating cell. While one can use different load values and matching network circuits\cite{badawe_true_2016}, we choose a load impedance of 70 $\Omega$ for simplicity in implementing the microstrip-type feeding network. This feed network uses three different $\lambda/4$ impedance transformers to arrive at the impedance of a 50 $\Omega$ feed port from all 64 cells. The length and width of the copper traces are calculated through microstrip transmission line equations for Rogers Ro4003 with a substrate thickness of 0.813 mm. Finally, they are optimized using CST to achieve optimal gain and bandwidth.

To demonstrate the scalability of the proposed structure at different frequencies, we have reduced the dimensions of the structure in Fig.~2 to one-tenth of the original size, except for the substrate thickness and metal layer, which are 0.1 mm and 17 $\mu$m, respectively. A wave-port with an impedance of 50 $\Omega$ was employed for the excitation of TMA from 50 GHz to 59 GHz. The simulation results for realized gain, directivity, radiation pattern, and $S_{11}$ at frequencies around 54.5 GHz demonstrate the TMA's ability to operate effectively at various frequencies (Fig.~12). The simulation results show that increasing the number of meta-atoms to 256 can achieve a realized gain, directivity, and radiation efficiency of about 19.6 dBi, 20.9 dBi and 73.5$\%$ at a frequency of 54.5 GHz. In addition, the simulation results show that the proposed TMA array, consisting of 16 strip lines, achieves a realized gain, directivity, and radiation efficiency
of approximately 19.5 dBi, 20.9 dBi and 73.2 $\%$, respectively, at a frequency of 54.5 GHz.

\begin{figure}[!t]
    \centering
    \hspace{10mm} 
    \subfloat[]{%
        \includegraphics[width=0.3\linewidth]{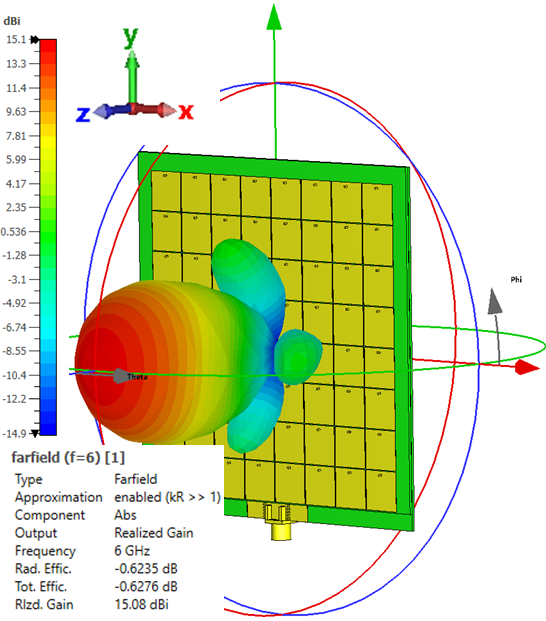}
    }
    \subfloat[]{%
        \includegraphics[width=0.3\linewidth]{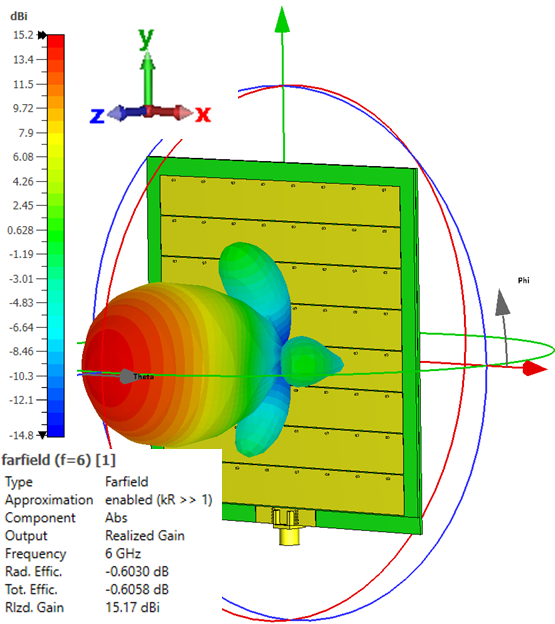}
    }
    \subfloat[]{%
        \includegraphics[width=0.35\linewidth]{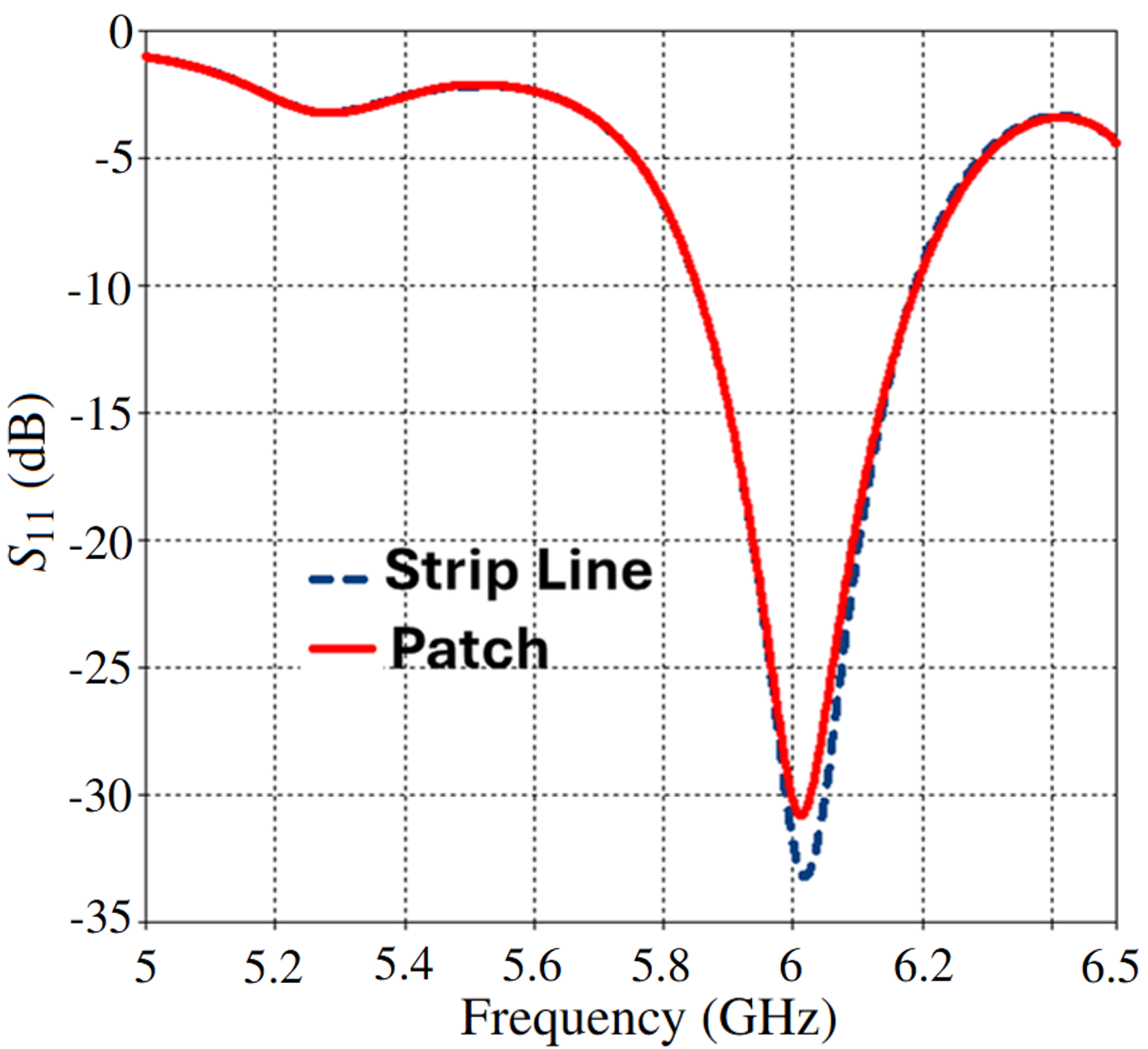}
   }
    \caption{The simulated 3D gain radiation pattern of the TMA with (a) squared-patch and (b) strip-line elements and (c) comparison between the $S_{11}$ of both array (a) and (b). [Note that in these simulation results, the feeding network and overall configuration of both arrays are identical, except that each row of patches in (a) is replaced by strip-lines in (b)]}
\end{figure}

\begin{figure}[!t]
    \centering
    \subfloat{%
        \includegraphics[width=0.45\linewidth]{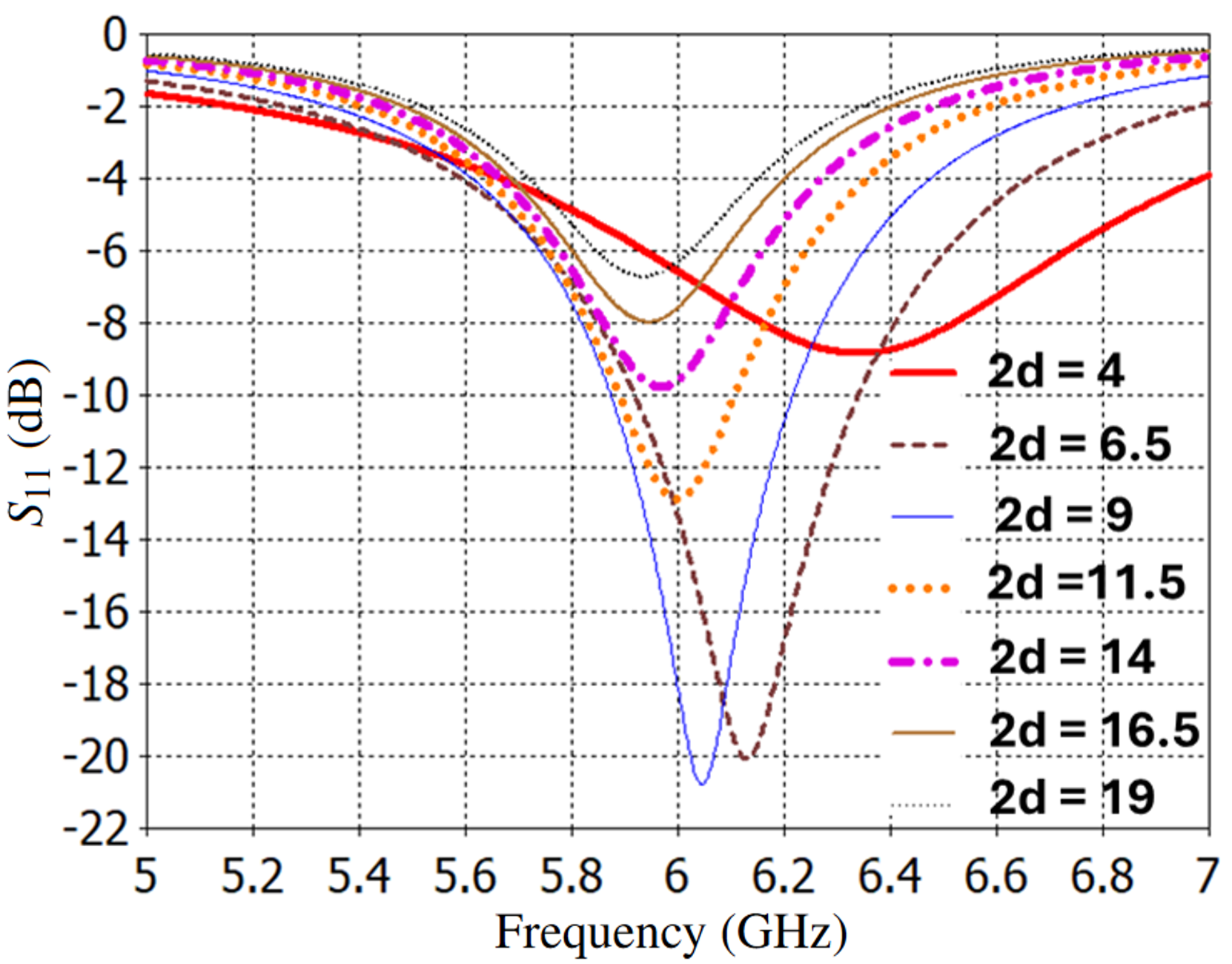}
    }
      \caption{Dependence of $S_{11}$ on inter-via distance along the x-axis [$2 \times d$ (mm)] when a normal incident plane wave illuminates an infinite strip-line array antenna. [Note that all dimensions of the strip-line cell are equal to the patch-cell dimensions described in Fig.~2, except that only two vertical slots are metallized and inter-via distance along the x-axis is varied between $d$ = 2 mm to $d$ = 9.5 mm]}

\end{figure}

\begin{figure}[!t]
    \centering
    \subfloat[]{%
        \includegraphics[width=0.40\linewidth]{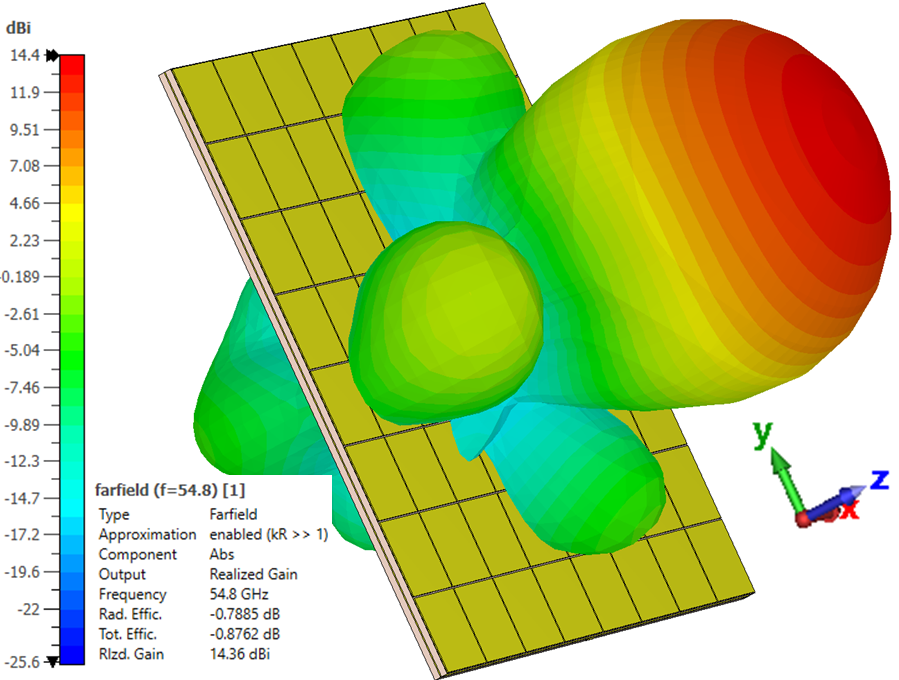}
    }
   \hspace{0.05\linewidth}
    \subfloat[]{%
        \includegraphics[width=0.40\linewidth]{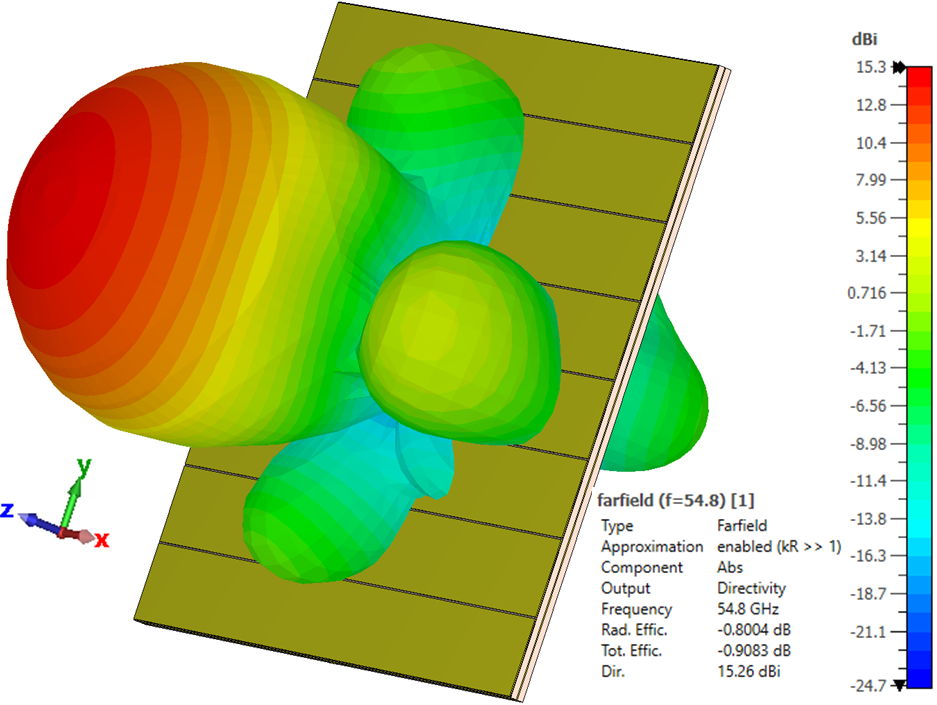}
    }
    \newline
    \subfloat[]{%
        \includegraphics[width=0.39\linewidth]{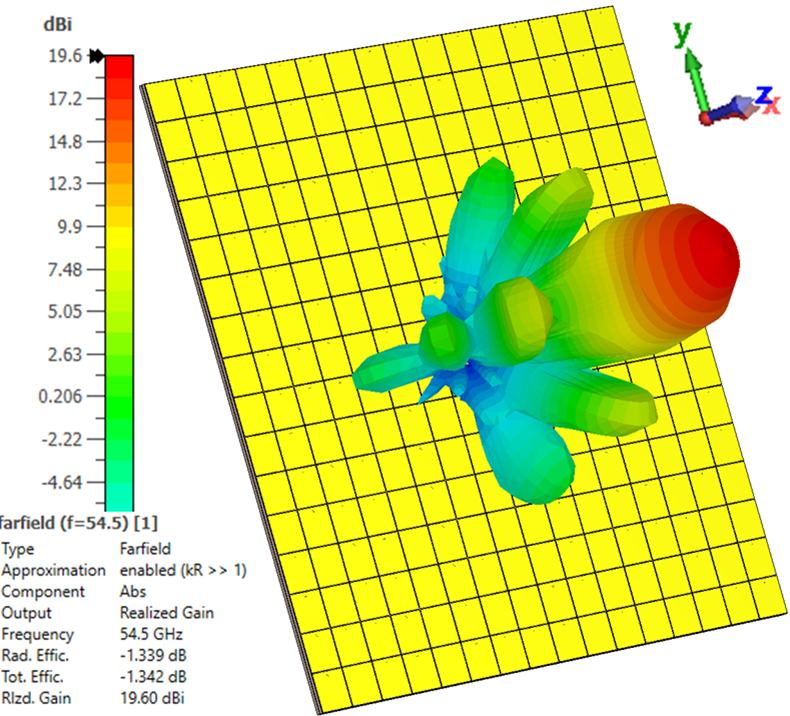}
    }
     \hspace{0.05\linewidth}
    \subfloat[]{%
        \includegraphics[width=0.40\linewidth]{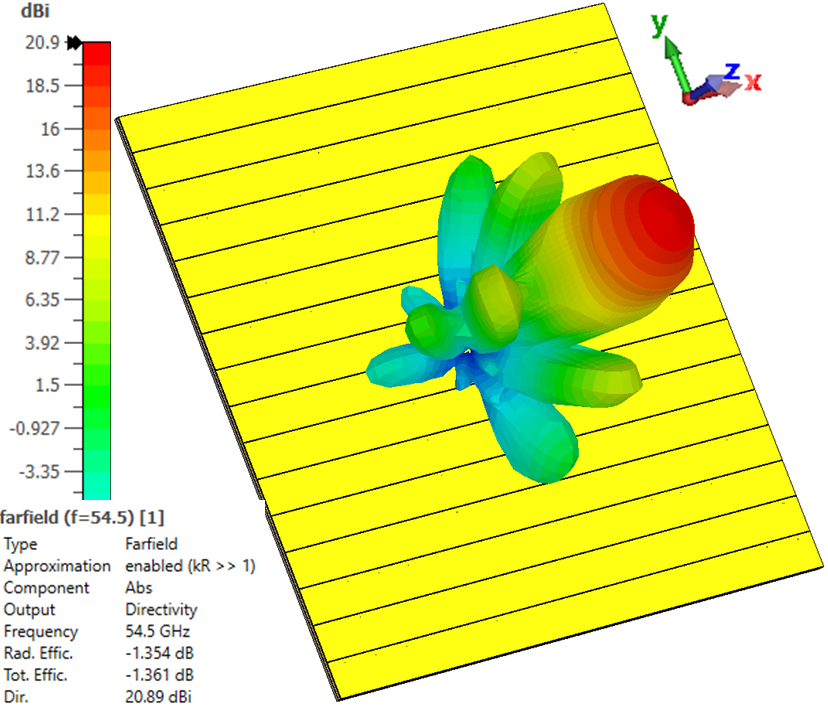}
    }
   \hspace{0.05\linewidth}
    \newline
    \subfloat[]{%
        \includegraphics[width=0.41\linewidth]{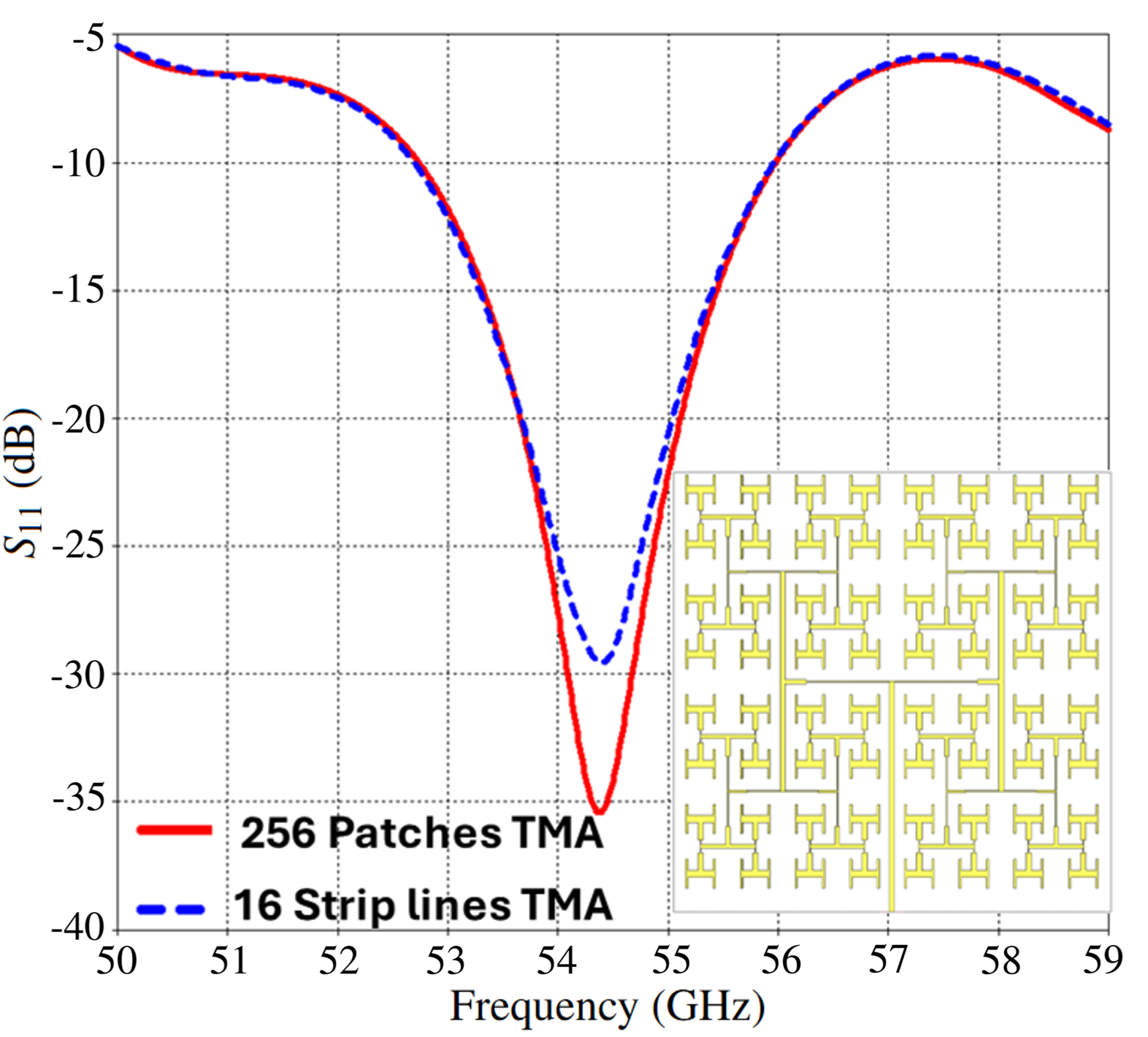}
    }
   
    \caption{ Simulated 3D directivity and gain radiation patterns at the resonance frequency of (a) 64 patch cells TMA (realized gain), (b) 8 striplines TMA (directivity), (c) 256 patch cells TMA (realized gain) and (d) 16 striplines TMA (directivity), and (e) $S_{11}$ of the TMA with 256 patch cells TMA and 16 strip lines TMA. [Noted that all dimensions of the radiating cell in Fig.~2 are scaled down by a factor of 0.1, except for the substrate thickness and copper layer, which are 0.1 mm and 17 $\mu$m, respectively.]}
\end{figure}

\section*{Experimental results and discussion}

A prototype antenna was produced using low-cost printed-circuit-board (PCB) technology based on the simulation TMA to validate the proposed design concept. Top and bottom views of the fabricated antenna prototype, with dimensions of 92.24 mm $\times$ 92.24 mm, are shown in Fig.~13. A milling machine (LPKF Protomat C60 Mill/Drill Unit Circuit Board Plotter) was employed to realize the feeding network pattern, patch, and shared ground structures. The shared ground metal layer of the first and second substrates was kept entirely metallic, and the anti-pads around each via with a radius of 1 mm were removed from these layers. Each of the holes created on the layers was soldered through wires. The feed network pattern of the prototype was milled on a Rogers Ro4003 substrate with a thickness of 0.813 mm.

\begin{figure}[!t]
    \centering
    \hspace{+20mm}
    \subfloat[]{%
        \includegraphics[width=0.70\linewidth]{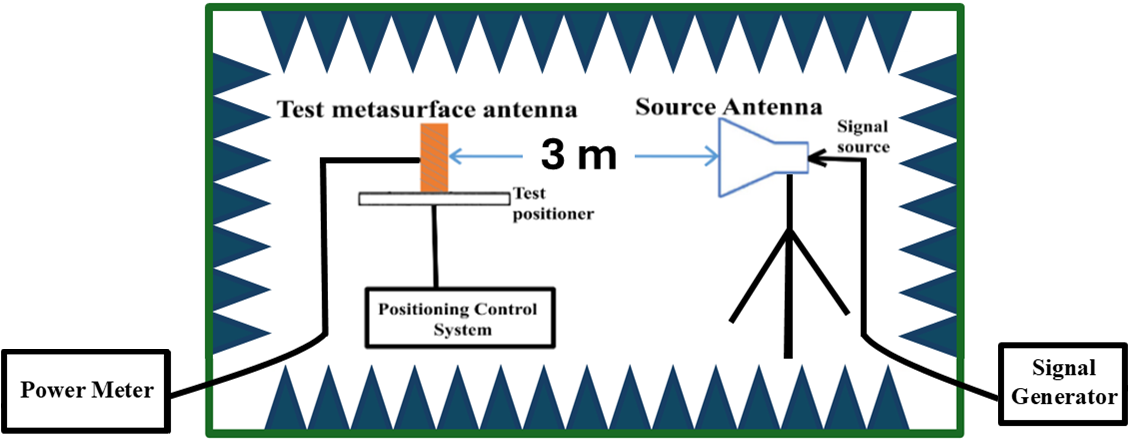}
    }
    \newline
    \subfloat[]{%
        \includegraphics[width=0.29\linewidth]{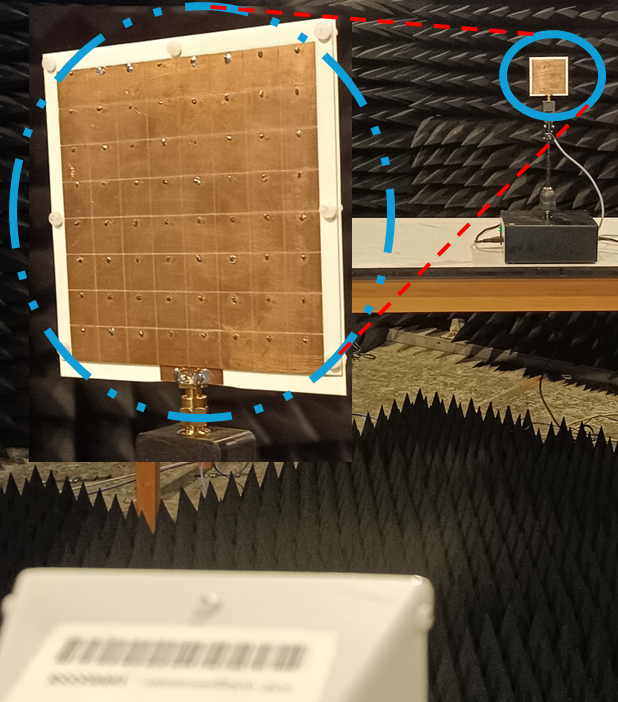}
    }
   \hspace{0.05\linewidth}
    \subfloat[]{%
    \includegraphics[width=0.31\linewidth]{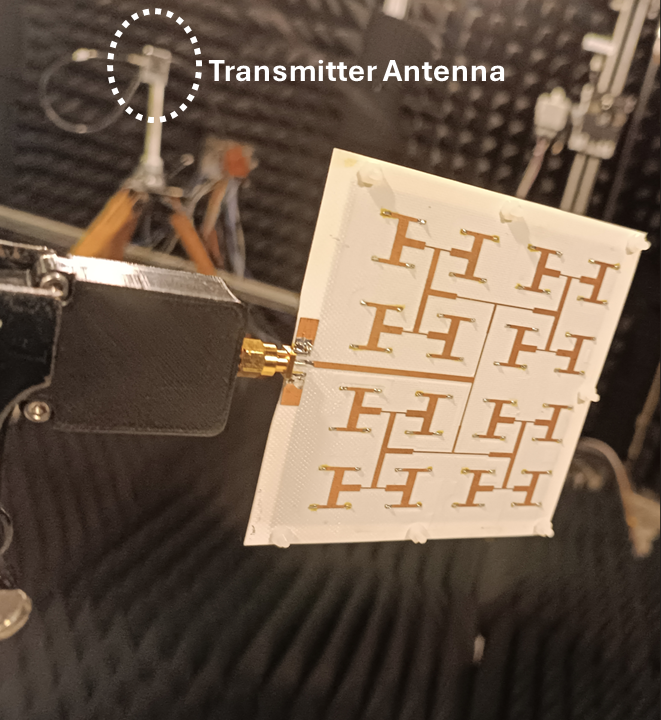}
    }
   
    \caption{(a) A schematic view of measurement setup in the anechoic chamber, (b) photograph of the front view of fabricated TMA and (c) back view of fabricated TMA.}
\end{figure}

\begin{figure}[!t]
    \centering
    \subfloat[]{%
        \includegraphics[width=0.3\linewidth]{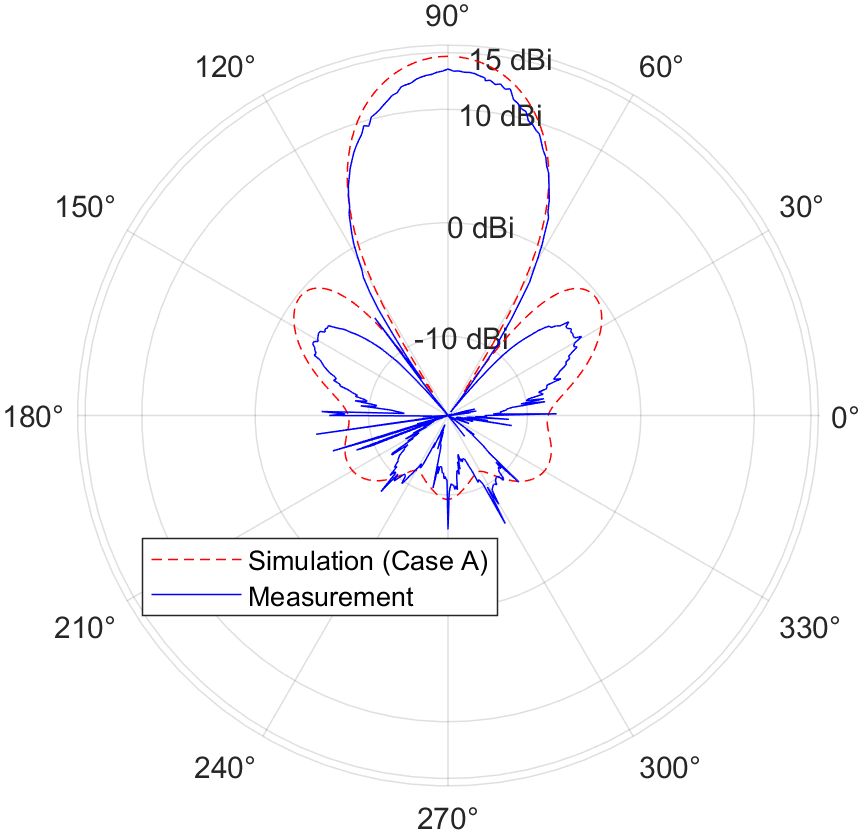}
    }
     \hspace{+5mm}
   \subfloat[]{%
        \includegraphics[width=0.44\linewidth]{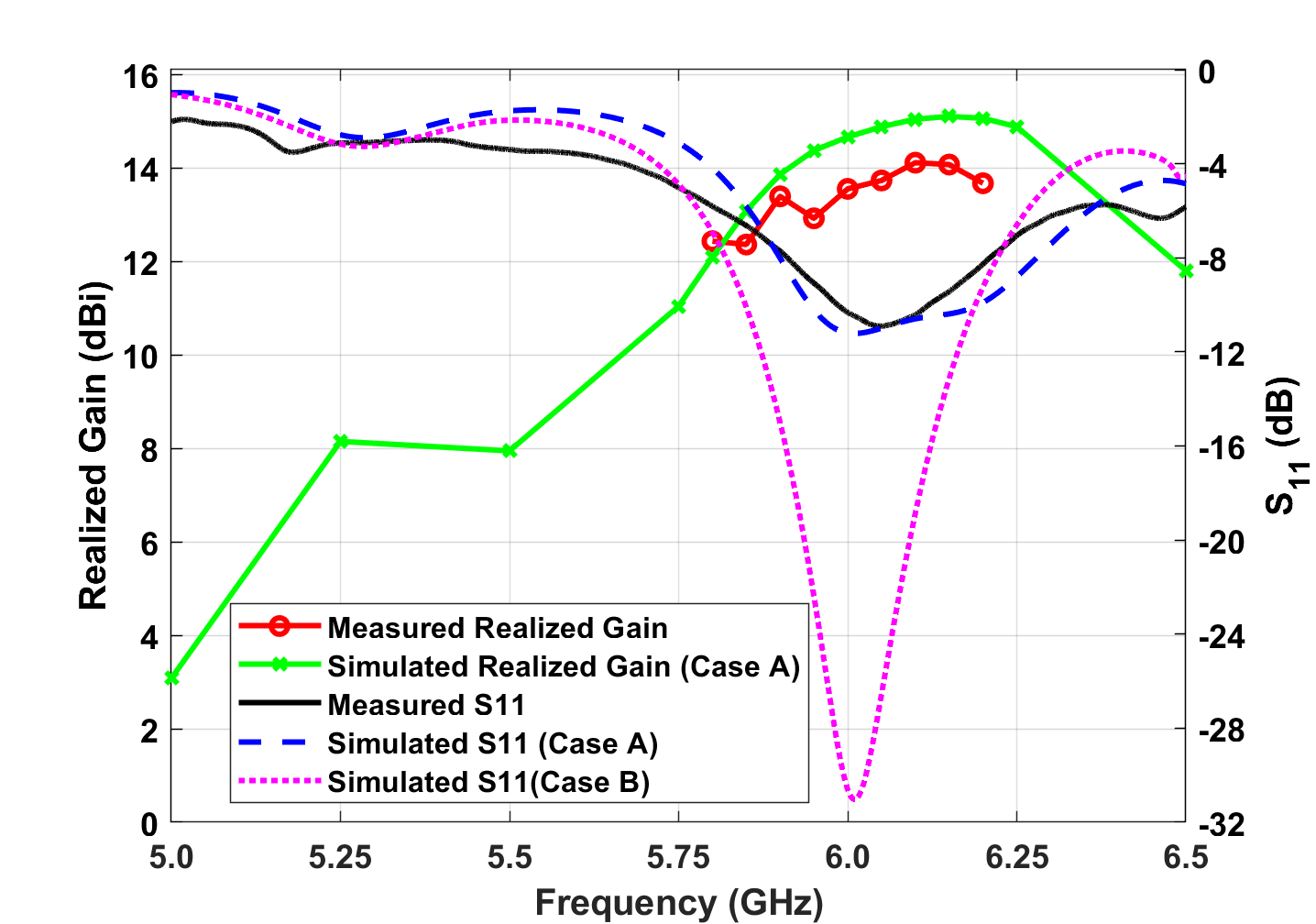}
    }
\caption{(a)  Simulated (dashed line) and measured (solid line) results of the normalized 2D far-field patterns of the designed TMA at the frequency of 6.0 GHz, and
(b) measured and simulated results of the realized gain and the $S_{11}$ of the designed TMA.
[Note: In the simulation results, Case A refers to removing 70 μm from the top surface of the substrate in the slots, while Case B refers to no removal from the thickness of top surface in the slots.]}

\end{figure}


During the production of the designed TMA using our PCB machine, a minor portion of the dielectric substrate layer was inadvertently removed along with the metal layer, reducing the substrate thickness in the slots on the top surface of the TMA. This reduction became a major source of discrepancy between the simulated and measured results.  In addition, fabrication imperfections in the soldering of the vias and SMA connector, along with losses in feeding patterns and measurement uncertainties and tolerances, further influenced the results \cite{ghaneizadeh_extremely_2020}. For a fair comparison between the simulated and measured results, we simulated the TMA with a 70 $\mu$m reduction in substrate thickness from the top surface in the slots, as shown in Case A in Fig.~14. For the sake of clarity, we included the simulation results of TMA without any reduction in thickness of its top surface, referred to as Case B in Fig.~14 and Table 1.  Additionally, it is important to note that an air gap may have formed between the top and bottom substrates, which could have affected the antenna’s performance, as the two layers were manually soldered and connected using vias or metal wires.

All 64 meta-radiators together are connected to a single 50 $\Omega$ SMA connector through a corporate isolated feed network. The Copper Mountain S5065 VNA was used to measure the $S_{11}$ at the feed point of the TMA. The comparison between the simulation and experimental results is shown in Fig.~14, indicating a good similarity between the experimental and simulated ones.

The radiation characteristics of the fabricated TMA were measured using a free-space test method in an anechoic chamber, as shown in Fig.~13. To determine the gain of our TMA, we performed relative gain measurements by comparing it with the known gain of a standard antenna. To do this, we employed a gain comparison method using two similar standard antennas (Schwarzbeck BBHA9210C antenna with a known gain) and our TMA. Initially, we measured the received power from the reference antenna (Schwarzbeck BBHA9210C) using a power meter. Next, we replaced the reference antenna with the TMA and recorded the received power using the same power meter. In both sets of measurements, all measurement elements maintained the same, except for the replacement of the receiving antennas. Additionally, the input power was kept constant throughout the experiments. Finally, we compared the received power of the TMA with that of the reference antenna and added the difference between them to the gain of the reference antenna. This approach for the antenna gain measurement is detailed in\cite{balanis2016antenna}. 
It should be noted that in this measurement, we assessed the realized gain of the TMA because the gain comparison method only involves replacing the receiving antenna. Consequently, the reflection and mismatch losses associated with the TMA's connection to the transmission line are inherently taken into account. Realized gain is defined as the product of reflection efficiency and gain as follows\cite{balanis2016antenna,Lynch_super_2024}:
\begin{equation}
\textit{Realized Gain} \triangleq \eta_{\text{r}} \times \textit{Gain}
\end{equation}
where $\eta_{\text{r}} = \text{reflection (mismatch) efficiency} = (1 - |S_{11}|^2)$. $\eta_{\text{r}}$ is dimensionless with $S_{11}$ the (voltage) reflection coefficient at the input terminal of the antenna \cite{balanis2016antenna}.
Another important parameter is the antenna’s radiation efficiency. It is defined as the ratio of gain to directivity, both of which are measured or calculated in the direction of maximum radiation\cite{balanis2016antenna}. Additionally, the aperture efficiency of the antenna\cite{balanis2016antenna} is given by 

\begin{equation}
\textit{Aperture Efficiency} = \textit{Realized Gain} \times \left( \frac{\lambda^2}{4\pi A} \right)
\end{equation}
where $\lambda$ denotes the free space wavelength, and $\textit{A}$ is the physical aperture area of the TMA, which we considered as an area consisting of 64 unit-cells, each of size $11.53 \times 11.53 \ \text{mm}^2$. In eq.~(6), it is assumed that the incoming wave is polarization-matched with the TMA.

It should be mentioned that the antenna was placed approximately 3 m from the standard horn antenna to ensure that the incident fields formed a plane wave \cite{badawe_true_2016}. A signal generator was connected to the standard horn antenna, and a power meter was employed to measure the received signal from the fabricated TMA as shown in Fig.~13. The linear polarization TMA was oriented so that the radiating electric field of the standard horn antenna was aligned parallel to the y-axis, as shown in Fig.~10. 
The measured and simulated antenna pattern results at a frequency of 6.0 GHz are shown in a polar format in Fig.~14(a) with the blue line and red dash lines, respectively. The compact configuration of the radiating cells reduces side lobes in the far-field radiation pattern. The measured antenna realized gain was conducted in the frequency range of 5.8 to 6.2 GHz with a 50 MHz step, achieving a realized gain of 13.5 (\(\pm\)0.5) dBi at 6.0 GHz. It should be mentioned that the accuracy of the realized gain measurements conducted within our anechoic chamber is approximately \(\pm\)0.5 dB. Fig.~14b shows the measured and simulated $S_{11}$ of the TMA. In addition, Fig.~14b shows the plot of the proposed TMA's measured and simulated realized gain. A comparison between the measured and simulated results shows good agreement.

\begin{table*}[!t]

\definecolor{Gray}{gray}{0.9}
\begin{center}
\begin{tabular}{p{2.8cm} p{1cm} p{1.3cm} p{1cm} p{2.3cm} p{1.2cm}p{2.3cm}}
\hline \hline
  \textbf{Ref.} &
  \textbf{Freq.\newline(GHz)} &
  \textbf{Gain (dBi)} &
   \textbf{$\varepsilon_r$ of\newline radiating cell}  
   & 
   \textbf{size ($\lambda_0$)} &  
    \textbf{Aperture\newline  efficiency}&  \textbf{ cell's shape} \\
\hline

  \cite{badawe_true_2016}(Sim.) &
  $2.97$ &
  $11.7$ &
  $9.9$ &
  $1.2\lambda_0\times1.2\lambda_0$  &
  80.6$\%$ &
  Cross strips \\
  
  \cite{badawe_true_2016}(Meas.) &
  $2.95$ &
  $9.4$ &
  $9.9$ &
  $1.2\lambda_0\times1.2\lambda_0$  &
  48.09$\%$ &
  Cross strips \\



\rowcolor{Gray}
 \cite{xu_integrated_2020}(Sim.) &
  $7.29$ &
  $19.7$ &
  $2.65$  &
   $3.1\lambda_0\times3.1\lambda_0$ &
  76.6$\%$&
  ELC \\

\rowcolor{Gray}
 \cite{xu_integrated_2020}(Meas.) &
  $7.29$ &
  $14.6$ &
  $2.65$  &
   $3.1\lambda_0\times3.1\lambda_0$ &
  24.3$\%$&
  ELC \\

   \cite{zhang_low_2024}(Sim.) &
  $5.8$ &
  $18.76$ &
  $2.65$  &
  *$3\lambda_0\times3\lambda_0$ &
  *65.66$\%$ &
  Ring$\&$patch \\

  \cite{zhang_low_2024}(Meas.) &
  $5.8$ &
  $17.76$ &
  $2.65$  &
  *$3\lambda_0\times3\lambda_0$ &
  *52.15$\%$ &
  Ring$\&$patch \\


  \rowcolor{Gray}
   \cite{fong_scalar_2010}(Meas.)&
  $8.7$ &
  $8.1$ &
  $n/a$  &
   $13.8\lambda_0\times13.8\lambda_0$ &
   $n/a$ &
  Patch\\

  \cite{Fan_pattern_2024}(Sim.) &
  $5.0$ &
  $21$ &
  $2.5$  &
  $4.4\lambda_0\times4.4\lambda_0$ &
  51.67$\%$ &
  Rect. patch\\

\rowcolor{Gray} 
  \cite{Abdelrahman_high_2014}(Sim.) &
  $11.3$ &
  $28.9$ &
  $2.574$  &
   $14.3\lambda_0\times14.3\lambda_0$ &
  30$\%$&
  Spiral-dipole \\

  \cite{koohestani_ultra_2021}(Sim.) &
  $0.868$ &
  $2.15$ &
   $n/a$ &
 $0.63\lambda_0\times0.63\lambda_0$ &
  32.2&
  Patch$\&$slots \\
  
   \rowcolor{Gray}
  \cite{cheng_w_2014}(Sim.) &
  $97.0$ &
  $28.81$ &
  $n/a$  &
  $30.9\lambda_0\times21.3\lambda_0$ &
  23$\%$&
  SIW-fed patch \\

   \cite{singh_microstrip_2019}(Sim.) &
  $5.83$ &
  $6.6$ &
  $4.4$  &
  $2.04\lambda_0\times1.55\lambda_0$ &
  11.45$\%$&
  Microstrip patch\\

  \rowcolor{Gray}
  Ours (Sim. Case B)&
  $6.0$ &
  $15.1$ &
  $3.26$  &
   $1.84\lambda_0\times1.84\lambda_0$ &
  75.3$\%$ &
  Patch\\

  Ours (Sim. Case B)&
  $6.0$ &
  $15.2$ &
  $3.26$  &
   $1.84\lambda_0\times1.84\lambda_0$ &
  76.8$\%$ &
  Strip-line\\

  \rowcolor{Gray}
   Ours (Sim. Case B)&
  $54.8$ &
  $14.4$ &
  $3.26$  &
  $1.68\lambda_0\times1.68\lambda_0$ &
  76.4$\%$&
  Patch \\

  Ours (Meas.)&
  $6.0$ &
  $13.5$(\(\pm\)0.5) &
  $3.26$  &
  $1.84\lambda_0\times1.84\lambda_0$ &
  52.9$\%$&
  Patch\\
  
 \hline
  \end{tabular}
  \end{center}
  \label{tbl:table1}
 \captionsetup{justification=raggedright,skip=5pt} 
 \caption*{*Aperture efficiency is calculated based on the gain and dimension of the array available in \cite{zhang_low_2024}.}
  \caption{Performance comparison to other published works}
\end{table*}



\textit{State-of-the-art metasurface designs.} Table 1 summarizes the performance of various published works involving different shapes of metasurface cells, normalized dimension sizes related to the operating wavelength, and dielectric constants. The proposed TMA demonstrates an aperture efficiency of 75.3$\%$ and 76.8$\%$ for square patch and strip-line (Case B) types, respectively, which is better than most other simulation results except in\cite{badawe_true_2016}. Although the design in\cite{badawe_true_2016} has a smaller dimension size, the dielectric constant used for the radiating cells in\cite{badawe_true_2016} is nearly three times that employed in our design. Furthermore, the dielectric layer thickness of their radiating unit-cells is 1.5 mm\cite{badawe_true_2016}, which is larger than that of the proposed TMA. Due to the limited number of published works in the field of TMA\cite{Yang_advanced_2024}, other antennas that use metasurfaces to enhance radiation performance are included in Table 1 for a fair comparison. For example, the proposed structure in \cite{fong_scalar_2010} and\cite{Abdelrahman_high_2014} uses a metasurface as a reflector and transmit array, respectively. The most recently published work\cite{Fan_pattern_2024} used a conventional rectangular patch array with a patch spacing of approximately half a free-space wavelength. The structure in \cite{zhang_low_2024} is presented for energy harvesting and polarization sensor applications. The authors in \cite{singh_microstrip_2019} designed a conventional microstrip patch antenna and in \cite{koohestani_ultra_2021} proposed a patch antenna on the designed metasurface ground. The antenna in \cite{xu_integrated_2020} employs three other metasurface layers on the TMA for coding of beam steering. The antenna in\cite{cheng_w_2014} operates at 97 GHz using the SIW-fed method. The simulation results show that, although the normalized dimensions of our TMA are significantly smaller than those of the antenna in\cite{cheng_w_2014}, our proposed antenna achieves a larger aperture efficiency than\cite{cheng_w_2014}. In addition, self-powered programmable metasurfaces were designed and implemented in \cite{Mu_self_2024} and \cite{chang_tailless_2024} using rectangular and square patches, respectively. However, in these designs, every four cells of the metasurface structures are connected through a power divider, functioning as a subarray. Therefore, it is not recommended to compare our single-port metasurface antenna with these structures, which have 25 output ports connected to 100 unit-cells.


\section*{Conclusion}
In summary, we have designed and experimentally demonstrated a metasurface antenna with high-aperture efficiency, using an HIS with vias located at the edge of the unit cell. This design does not require additional resonant elements or air gaps. The prototype comprises an 8$\times$8 array of patch resonators and a 64-way feeding network manufactured using standard PCB technology. We developed a circuit model for this structure and verified it through full-wave simulations. Subsequently, we simplified the design by interconnecting the vertical slots of each square patch, transforming the patches into parallel strip lines and introducing a new antenna configuration. 
The fabricated TMA achieved a realized gain of 13.5(\(\pm\)0.5) dBi at 6.0 GHz and a (higher) peak realized gain of 14.1(\(\pm\)0.5) dBi at 6.1 GHz. The measured outcomes are well consistent with the simulated outcomes. Furthermore, the scalability of the design (Case B) was verified through full-wave simulations, demonstrating that scaling down the patch cells and reducing the substrate thickness can increase the resonance frequency to 54.8 GHz, achieving an aperture efficiency of 76.4$\%$ and realized gain of 14.4 dBi and bandwidth from 53.36 GHz to 56.34 GHz. This scalability presents potential applications across a broad frequency range, from megahertz to sub-terahertz spectra, and offers a versatile platform for future advancements in 6G wireless communication systems.

\bibliography{sample}

\section*{Acknowledgements }

This work has been funded by the Deutsche Forschungsgemeinschaft (DFG, German Research Foundation)-grant number 511400365, JO 1413/3-1. The authors would like to thank Mr.~Uwe Pagel for his technical support during the fabrication process and Mr.~Fred Becker for his technical support in the antenna measurement.

\section*{Funding}
Deutsche Forschungsgemeinschaft (DFG, German Research Foundation)-grant number
511400365, JO 1413/3-1.

\section*{Author contributions statement}

A.G.~and M.J.~conceived the idea, designed and fabricated the structure. A.G. carried out the electromagnetic simulations. A.G., S.P.~and M.S.~conducted the antenna measurements. All authors analyzed the results and reviewed the manuscript.

\section*{Competing interests}

The authors declare no competing interests.

\section*{Additional information}

Correspondence and requests for materials should be addressed to M.J.

\end{document}